\documentclass[aps,pra,superscriptaddress,reprint]{revtex4-1}
\usepackage[colorlinks=true,citecolor=blue,linkcolor=magenta]{hyperref}
\usepackage{graphicx}
\usepackage{bm}
\usepackage{dsfont}
\usepackage{amsmath,amssymb}
\usepackage{color,psfrag}
\usepackage{pstool}
\usepackage{array}
\usepackage{longtable}
\usepackage{wrapfig}
\usepackage{bbm}
\usepackage{color}

\newcommand{\bra}[1]{\langle #1|}
\newcommand{\ket}[1]{|#1\rangle}
\newcommand{\ex}[1]{\langle #1 \rangle}

\begin{document}

\author{Andreas Albrecht}
\affiliation{Institut f\"ur Theoretische Physik, Albert-Einstein-Allee 11, Universit\"at Ulm, 89069 Ulm, Germany}
\affiliation{Center for Integrated Quantum Science and Technology, Universit\"at Ulm, 89069 Ulm, Germany}
\author{Guy Koplovitz}
\affiliation{Department of Applied Physics, The Hebrew University of Jerusalem, Givat Ram, Jerusalem 91904, Israel}
\affiliation{Institute of Chemistry, The Hebrew University of Jerusalem, Givat Ram, Jerusalem 91904, Israel}
\affiliation{The Harvey M. Krueger Family Center for Nanoscience and Nanotechnology, The Hebrew University of Jerusalem, Jerusalem 91904, Israel}
\author{Alex Retzker}
\affiliation{Racah Institute of Physics, The Hebrew University of Jerusalem, Jerusalem 91904, Israel}
\affiliation{The Harvey M. Krueger Family Center for Nanoscience and Nanotechnology, The Hebrew University of Jerusalem, Jerusalem 91904, Israel}
\author{Fedor Jelezko}
\affiliation{Institut f\"ur Quantenoptik, Albert-Einstein-Allee 11, Universit\"at Ulm, 89069 Ulm, Germany}
\affiliation{Center for Integrated Quantum Science and Technology, Universit\"at Ulm, 89069 Ulm, Germany}
\author{Shira Yochelis}
\affiliation{Department of Applied Physics, The Hebrew University of Jerusalem, Givat Ram, Jerusalem 91904, Israel}
\affiliation{The Harvey M. Krueger Family Center for Nanoscience and Nanotechnology, The Hebrew University of Jerusalem, Jerusalem 91904, Israel}
\author{Danny Porath}
\affiliation{Institute of Chemistry, The Hebrew University of Jerusalem, Givat Ram, Jerusalem 91904, Israel}
\affiliation{The Harvey M. Krueger Family Center for Nanoscience and Nanotechnology, The Hebrew University of Jerusalem, Jerusalem 91904, Israel}
\author{Yuval Nevo}
\affiliation{The Robert H. Smith Institute of Plant Sciences and Genetics in Agriculture, The Hebrew University of Jerusalem, Rehovot 76100, Israel}
\affiliation{The Harvey M. Krueger Family Center for Nanoscience and Nanotechnology, The Hebrew University of Jerusalem, Jerusalem 91904, Israel}
\author{Oded Shoseyov}
\affiliation{The Robert H. Smith Institute of Plant Sciences and Genetics in Agriculture, The Hebrew University of Jerusalem, Rehovot 76100, Israel}
\affiliation{The Harvey M. Krueger Family Center for Nanoscience and Nanotechnology, The Hebrew University of Jerusalem, Jerusalem 91904, Israel}
\author{Yossi Paltiel}
\affiliation{Department of Applied Physics, The Hebrew University of Jerusalem, Givat Ram, Jerusalem 91904, Israel}
\affiliation{The Harvey M. Krueger Family Center for Nanoscience and Nanotechnology, The Hebrew University of Jerusalem, Jerusalem 91904, Israel}
\author{Martin B Plenio}
\affiliation{Institut f\"ur Theoretische Physik, Albert-Einstein-Allee 11, Universit\"at Ulm, 89069 Ulm, Germany}
\affiliation{Center for Integrated Quantum Science and Technology, Universit\"at Ulm, 89069 Ulm, Germany}
\title{Self-assembling hybrid diamond-biological quantum devices}

\maketitle

\textbf{The realization of scalable arrangements of nitrogen vacancy (NV) centers in diamond remains a key challenge on the way towards efficient quantum information processing, quantum simulation and quantum sensing applications. Although technologies based on implanting NV-center in bulk diamond crystals\,\citep{pezzagna11} or hybrid device approaches\,\citep{santori10} have been developed, they are limited in the achievable spatial resolution and by the intricate technological complexities involved in achieving scalability. We propose and demonstrate a novel approach for creating an arrangement of NV-centers, based on the self-assembling capabilities of biological systems and its beneficial nanometer spatial resolution\,\citep{aldaye11, behrens08}. Here, a self-assembled protein structure serves as a structural scaffold for surface functionalized nanodiamonds, in this way allowing for the controlled creation of NV-structures on the nanoscale and providing a new avenue towards bridging the bio-nano interface. One-, two- as well as three-dimensional structures\,\citep{douglas09} are within the scope of biological structural assembling techniques. We realized experimentally the formation of regular structures by interconnecting nanodiamonds using biological protein scaffolds. Based on the achievable NV-center distances of 11nm, we evaluate the expected dipolar coupling interaction with neighboring NV-center as well as the expected decoherence time. Moreover, by exploiting these couplings, we provide a detailed theoretical analysis on the viability of multiqubit quantum operations, suggest the possibility of individual addressing based on the random distribution of the NV intrinsic symmetry axes and address the challenges posed by decoherence and imperfect couplings.  We then demonstrate in the last part that our scheme allows for the high-fidelity creation of entanglement, cluster states and quantum simulation applications.}

\begin{figure*}[tb]
\begin{centering}
\includegraphics[scale=0.38]{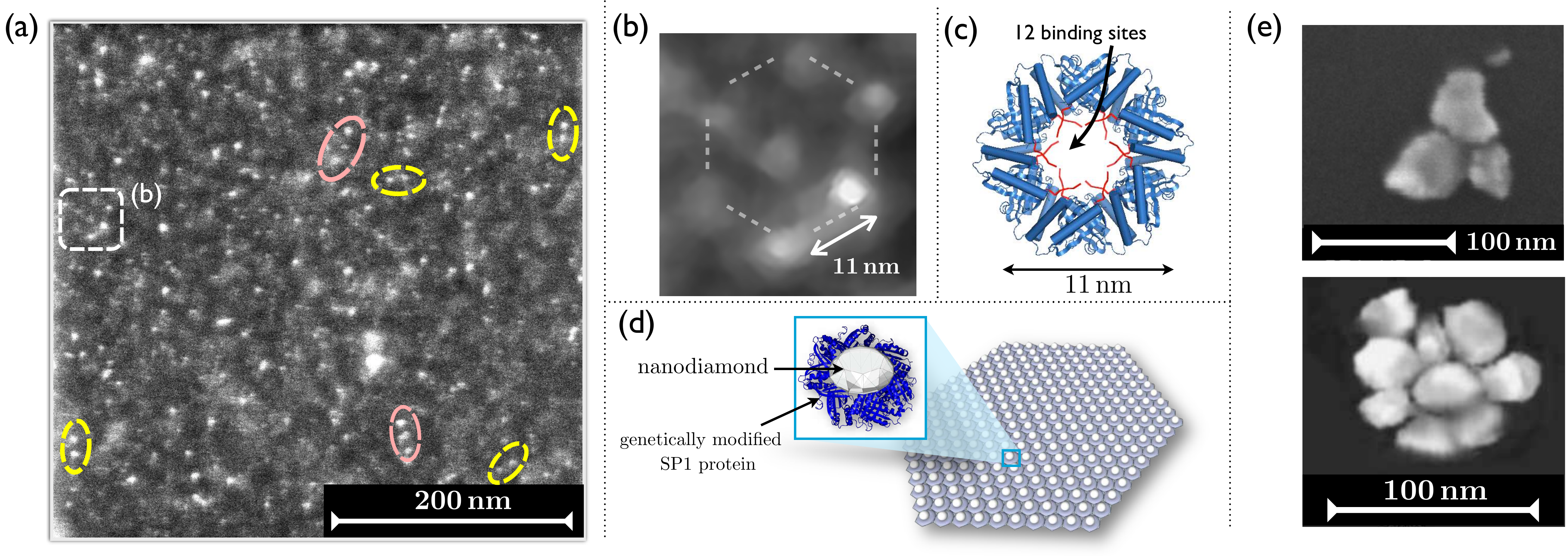}
\caption{\label{b_NDarray} \textbf{Nanodiamond (ND)-SP1 arrays and clusters} \textbf{(a)} DF-STEM (Dark field scanning transmission electron microscopy) image of ND structures on an SP1-ordered monolayer (ND diameter\,$\sim$5nm). The hexagonal arrangement in the white dashed square is magnified in part (b). Yellow and red circles show diamond dimers and trimers, respectively, with inner distances of 11\,nm. \textbf{(b)} Enlarged section of the white dashed square of (a) showing a hexagonal structure formed of 7 NDs. The symmetry and distances are determined by the underlying SP1-layer. \textbf{(c)} SP1-protein ring: The inner linkers (binding sites) are genetically modified to enable graphite specific binding. \textbf{(d)} Schematic of an ordered hexagonal array of SP1-NDs hybrids consisting of a ND attached to the SP1 inner cavity. Here the SP1-monolayer serves as a structural scaffold. \textbf{(e)} SEM image of larger (ND diameter\,$\sim$30nm) clusters connected by SP1 and obtained in solution.}
\end{centering}
\end{figure*}

\begin{figure}[htb]
\begin{centering}
\includegraphics[scale=0.35]{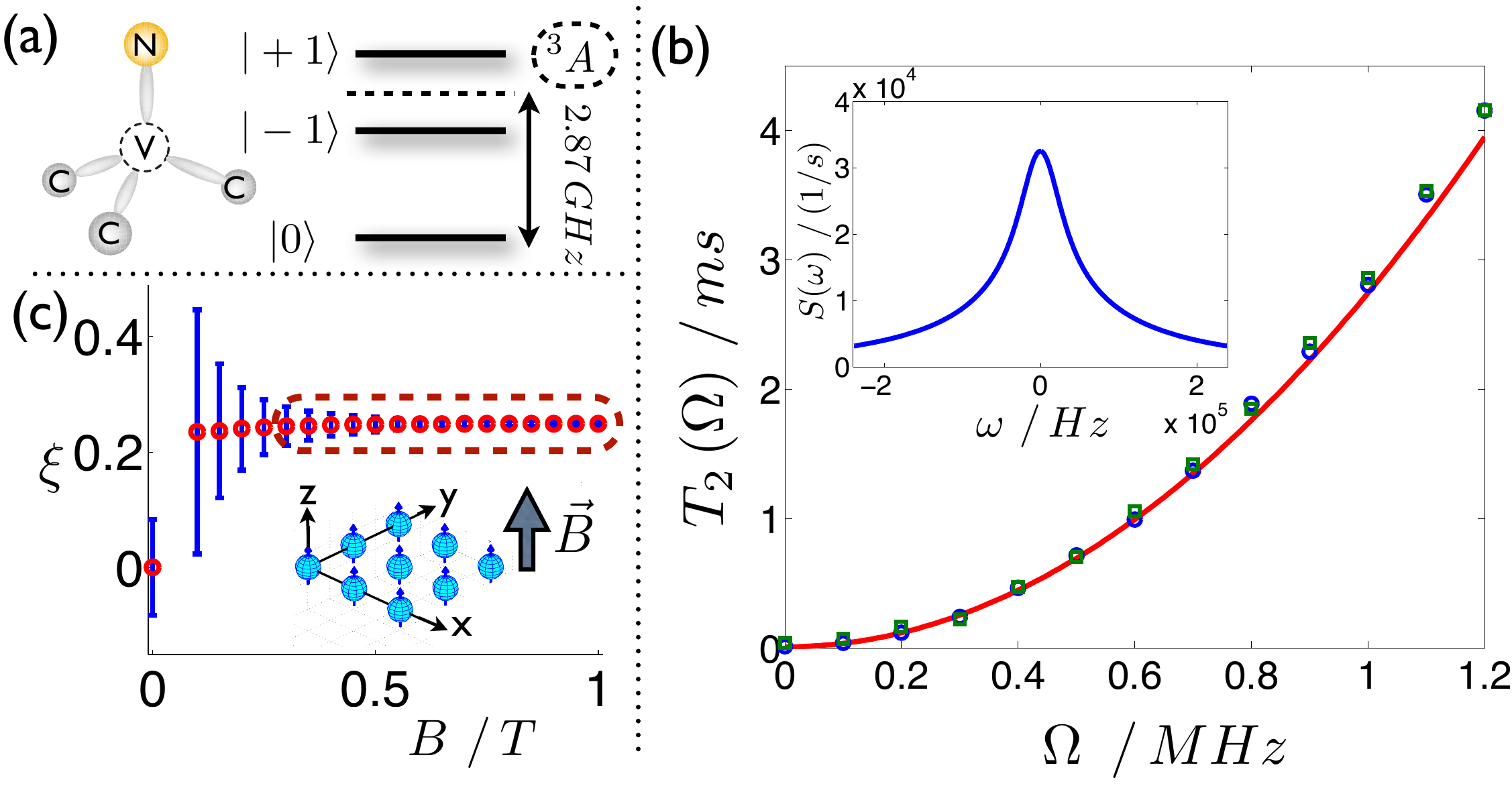}
\caption{\label{b_dynamdecoupl} \textbf{NV center, dynamical decoupling and dipolar coupling strength} \textbf{(a)} Ground state electron spin triplet of the nitrogen vacancy center with zero field splitting $2.87\,{\rm GHz}$ and the degeneracy of the $\ket{\pm 1}$ states lifted by a weak magnetic field. \textbf{(b)} Effective $T_2$ time obtained by dynamical decoupling vs strength of the decoupling field: The continuous line corresponds to formula\,(\ref{t2eff}), circles correspond to the 1/e decay time and squares to an exponential fit both obtained by a numerical noise simulation. Noise parameters follow from the noise spectrum (inset) calculated using the method provided in\,\citep{sousa09}: $\tau=2.5\,\mu s$, $b=30.2\,{\rm kHz}$, $T_2(\Omega=0)=13.3\,\mu s$. \textbf{(c)}  Dipolar coupling parameter $\xi$ vs the external magnetic field. Red circles denote the average and blue lines the  variance arising from the random axis orientation. For large magnetic field the quantization axis is given by the external field and $\xi\simeq 1/4$. The green dashed area indicates the range that can be corrected using compensation methods. All parameters are obtained by averaging over $10^6$ random spin orientations. Inset: Optimal magnetic field configuration for a 2D array as also used for the main graph.  }
\end{centering}
\end{figure}

Coupled solid-state spin systems as electron spins in quantum dots, phosphorous donors in silicon and color centers in diamond form promising candidates for the emerging field of quantum technologies\,\citep{awschalom13}. Among those the negatively charged nitrogen-vacancy center (NV$^-$) in diamond\,\citep{doherty13}, composed of a substitutional nitrogen atom and an adjacent vacancy within the diamond lattice, subject of this work,  stands out due to its long coherence time up to milliseconds at room temperature\,\citep{balasubramanian09}.  It forms a discrete atom-like energy level structure within the diamond bandgap, with the ground state described by an electron-spin triplet (Spin-1) as illustrated in figure\,\ref{b_dynamdecoupl}\,(a), that allows for the full coherent control by means of e.g. microwave drivings and static magnetic fields\,\citep{jelezko04}. Remarkably, readout and initialization through the excited state can be performed optically, taking benefit of the spin-dependent fluorescence rates and an intersystem crossing, the latter one allowing for the high-fidelity state preparation by optical pumping\,\citep{jelezko04}. Interaction among neighbouring NV-centers can be mediated by the magnetic dipolar coupling of the electronic spins\,\citep{dolde13, bermudez11}, yet this requires distances of the order of 10\,nm or below to enable coherent interaction strengths that comfortably exceed the decoherence times.  However current techniques for the controlled positioning of NV-centers within bulk crystals, i.e. the creation of NV-centers by ion implantation, are limited to several tens of nanometers in position\,\citep{staudacher12}.  
In contrast, the ability of biological systems for structural self-assembly\,\citep{aldaye11, behrens08} is a powerful tool allowing for the simple and parallel creation of large ordered arrays on nanometer scales, that holds the potential for outperforming the limited resolution and structural complexity achievable of conventional lithography based on serial pattern creation. We propose to combine the self-assembly of biological systems with the guided attachment of surface functionalized nanodiamonds. The biomolecules are used as a structural scaffold to enable the formation of NV-center configurations with high spatial resolution. This can be achieved with tiled motifs such as short DNA strands\,\citep{seeman07} or membrane forming complexes including SP1\,\citep{medalsy08}, LH1\,\citep{kondo07} and TF55$\beta$\,\citep{mcmillan05}, that allow for the creation of one and two dimensional arrays. Going beyond two-dimensional periodic patterns, the method of DNA-origami\,\citep{rothemund06, douglas09}, based on folding a large single stranded DNA molecule directed by staple strands, enables the creation of more complex highly controllable structures, ranging from aperiodic arrays to real three dimensional  structures. Elaborate knowledge in genetic engineering now allows control over the assembly, structure and topology of biomolecular complexes as well as the attachment of nanoparticles with nanometer precision \,\citep{kuzyk12}.
Each of these structures is suitable for hosting nanodiamonds, whose size can be controlled down to 4\,nm in fabrication\,\citep{tisler09}, and whose chemical attachment and site-specific binding can be directed by surface functionalization and labeling. This allows for the high precision positioning of NV-centers incorporated in such diamonds, which subsequently interact via dipolar couplings, and paves the way for a highly controllable array of interacting quantum systems.

\textbf{Formation of SP1 nanodiamond structures} --- To demonstrate the feasibility of interconnecting nanodiamonds\,(NDs) with biological structures, we present the controllable and repeatable formation of small nanodiamond complexes using an SP1 (Stable Protein\,1, see figure\,\ref{b_NDarray}(c) and Supplement section\,1) protein variant and first steps towards the formation of regular arrays of NDs on SP1 arrays. A crucial first step to enable both experiments is the genetic modification of SP1 to fuse 12 graphite-specific binding peptides to the SP1 N-terminus, that permit site specific binding of SP1 to the carbon sp2-hybridization which forms on the nanodiamond surface\,\citep{wolf12}.

Site-specific binding of the SP1 is essential for small NDs (under 10 nm) to form regular structures on an SP1 array (figure\,\ref{b_NDarray}\,d)\,\citep{guy13}.  As an important result towards the creation of large ordered NDs structures, we achieved the formation of numerous dimers and trimers along with larger ordered structures such as a 7\,NDs hexagon as illustrated in figure\,\ref{b_NDarray}\,(a)\&(b).  Here a monolayer of genetically modified SP1 was formed by the Langmuir-Blodgett method\,\citep{heyman09ii} and subsequently combined with a ND solution (5nm in diameter formed by laser ablation). Using both the SP1-template and a diluted ND solution, ordering of the NDs partially filling the SP1 template has been achieved. Such isolated structures are promising candidates for NV-coupling experiments. Figure\,\ref{b_NDarray}\,(a) shows an image of ND structures on the SP1 monolayer and figure\,\ref{b_NDarray}\,(b) is enlarging the ND hexagon. In this SP1/NDs sample, the size of the measured nano-particles is around 5\,nm in diameter, therefore matching the expected NDs size. On the other hand, the distance between centers of adjacent particles corresponds to 11nm, which is the distance between the SP1 proteins on an ordered layer. Moreover, electron diffraction on this sample in the relevant areas proved the existence of diamonds on the surface\,(see Supplementary section\,1).

Two references were used to ensure that the combination of the SP1 template and the NDs is essential for achieving the uniform spacing between adjacent NDs. A sample without the SP1 template that contains only adsorbed NDs, in which case only irregular ND aggregates were observed in the TEM measurements, and a sample with the SP1 template but without the NDs in which case we did not find any nano-particles on the sample.

In a different approach, small ND structures were also achieved using larger (average diameter 30\,nm created by grinding) NDs.  The ND clusters were formed by mixing solutions of NDs and SP1 under ambient conditions. The average number of NDs in such a complex can be controlled by the concentration ratio of SP1 to NDs. Mixing a 1mg/ml ND solution with a 1mg/ml SP1 solution at a ratio of 1:1 mainly leads to the formation of dimers and trimers and for even higher concentrations of SP1 large diamond clusters can be observed (figure\,\ref{b_NDarray}\,(e)). As the NDs are larger than the SP1, the exact structure of the clusters is controlled by the nanodiamond shape as several SP1 will bind the NDs across a surface. To verify that the creation of the clusters is due to the binding with SP1 rather than electrostatic forces we have also followed the same procedure in the absence of SP1. In this case no ND clusters are observed.

\textbf{Decoherence and dynamical decoupling} --
A key challenge for quantum computation and simulation with self-assembled arrays of nanodiamonds are the currently achieved relatively short coherence times of a few $\mu s$\,\citep{tisler09} compared to the typical coupling strength of several kHz between NV-centers  in different nanodiamonds\,\citep{neumann10, bermudez11}. This is originated in interactions of the NV-center with the very proximate surface spins\,\citep{hall10} and external charges, such that substantial improvements can be obtained by means of surface functionalization methods\,\citep{jianming12, tisler09}. The remaining decoherence after optimization of material design can be further improved by applying dynamical decoupling techniques, that allow to increase the coherence time by several orders of magnitude.
 These may be implemented either as pulsed schemes\,\citep{lange2010} or by continuous driving fields\,\citep{bermudez11, timoney11, cai11,cai13}. At this point we would like to point out that beside surface spins, other impurities well-known from bulk diamond such as spins of C-13 (type-IIa diamond) and P1-centers of nitrogen donors (type-Ib diamond), contribute to the effect of decoherence as well. This influence however is much less dominant as their low concentration leads to a rather small and only weakly coupling spin bath\,\citep{tisler09}. Moreover it can be described and thus decoupled in the same framework as outlined below and consequently we do not expect those additional impurities to change the results presented here.  \\The pure dephasing noise can be modelled as a fluctuating magnetic field energy shift $b(t)$  with zero mean and $\langle b(t)\,b(0) \rangle = b^2\,e^{-t/\tau}$\,(see Supplement section\,5). Adding a continuous driving interaction for decoupling, i.e. a resonant microwave coupling $\Omega$ on the relevant electron spin transition, the Hamiltonian of the effective two level system in a frame rotating with the laser frequency for a single NV-center is given by $H=(b(t)/2)\,\sigma_z+(\Omega/2)\,\sigma_x$,
where $\sigma_{x,z}$ are the usual spin-1/2 Pauli operators.
In the relevant Markovian limit $t\gg\tau$ and $\Omega\,t>1$ the decoherence decay rate $R(t)$ is determined by the noise spectrum $S(\omega)$ evaluated at the decoupling frequency, i.e. $R(t)\propto S(\Omega)$ and the more general formalism will be discussed in the Supplement\,(section\,2)\,\citep{gordon07}. This allows to define an effective $T_2$-time of the system resulting in
\begin{equation}\label{t2eff} T_2(\Omega)=\dfrac{1+\Omega^2\,\tau^2}{b^2\,\tau}=T_2(\Omega=0)\,\left( 1+\Omega^2\,\tau^2 \right) \end{equation}
for the considered Lorentzian noise spectrum.
Figure\,\ref{b_dynamdecoupl}\,(b) compares this formula to numerical noise simulations showing that both the $T_2$-scaling and the exponential decay behaviour are in good agreement with numerical calculations. Here the noise spectrum was calculated for a fluorine terminated surface of a spherical diamond with radius $r=5\,{\rm nm}$ that results in a fluorine nuclear spin-1/2 surface with a nearest neighbour distance of $2.5\,$\r{A}, using the mean-field approach described in\,\citep{sousa09}\,(see Supplement, section\,4). We thus expect a noise correlation time of $\tau=2.5\mu s$ and a mean square root amplitude $b=30.2\,{\rm kHz}$ ($\tau\propto 1/(n^{3/2})$ and $b^2\propto n^2\,r$ with $n$ the surface spin density and $r$ the nanodiamond radius), that could be further improved by spin bath polarization and decoupling schemes\,\citep{jianming12, goldburg63}. For those parameters, a decoupling field of $\Omega=1.2\,{\rm MHz}$ leads to an effective coherence time of $T_2\simeq 4\,{\rm ms}$ comparable to the electron spin $T_1$-time that imposes an upper limit on the dynamical decoupling. In experiments it is possible to achieve much higher decoupling fields up to 300\,MHz such that even a small number of possible electron spins can be decoupled\,(for a discussion of electron spin noise see Supplement, section\,4).
Combining self-assembled structures with lithographic techniques might help to integrate current structures within the system\,\citep{penzo11} beneficial especially for high driving fields.
Importantly, decoherence due to intensity fluctuations of the decoupling field can be suppressed by using a concatenated decoupling scheme as proposed in\,\citep{cai11}.

\textbf{Engineering of interactions, spin gates and quantum simulation}\label{elspinsect} ---
Dipolar interactions between electron spins of two adjacent NV-centers provide a possibility for implementing gate operations\,\citep{dolde13, bermudez11}.
Combined with the continuous driving of the decoupling field, the total effective Hamiltonian in a frame rotating with the microwave frequency of the driving field can be written as\,(see Supplement, section\,3)
\begin{equation}\label{htot}  H'\simeq \sum_i \dfrac{b_i(t)}{2}\,\sigma_z^i + \sum_i\dfrac{\Omega_i}{2}\,\sigma_x^i+\sum_{i>j}\dfrac{J_{ij}}{2}\,\sigma_z^i\,\sigma_z^j\,. \end{equation}
Herein the first two parts describe the dephasing noise and decoupling for the individual NV-centers as introduced in the previous section, respectively.  The last part accounts for the dipolar coupling with $J_{ij}=2\,\xi_{ij}\mu_{ij}$, where $\xi_{ij}$ depends on the external magnetic field strength\,$|\vec{B}|$ and its orientation with respect to both the NV symmetry axes and the vector $\vec{r}_{ij}$ connecting NV-center $i$ and $j$. In the limit of high magnetic fields $\xi_{ij}=1/4\,(1-3\,\cos^2\theta_{ij})$, where $\theta_{ij}=\angle (\vec{r}_{ij},\vec{B})$ and $\mu_{ij}=\mu_0\gamma_{el}^2\hbar/(4\pi r_{ij}^3)$ with the latter being $\mu_{ij}=52\,{\rm kHz}$ for an NV-center distance  $r_{ij}=10\,{\rm nm}$. $\mu_0$ and $\gamma_{el}$ denote the magnetic permeability and the electron gyromagnetic ratio, respectively. It is important to note that the configuration of nanodiamonds leads to a random relative orientation of the NV symmetry axes in space, the latter forming a `natural' quantization axis along the N-V direction by the associated crystal-field splitting of the ground state triplet ($D\sim 2.87\,{\rm GHz}$) (see figure\,\ref{b_dynamdecoupl}(a)).  This leads to two important consequences: A uniform quantization axis has to be defined by a sufficiently strong external magnetic field to guarantee a uniform dipolar coupling, as illustrated in figure\,\ref{b_dynamdecoupl}\,(c) together with its optimal 2D-orientation.  As shown in the figure, this can be achieved applying a magnetic field $B\gtrsim 0.5\,{\rm T}$ ($\gamma_e\,B/(2\,\pi)\gtrsim 14\,{\rm GHz}$), in which case the magnetic energy shift dominates over the crystal-field splitting ($\gamma_e\,B\gg D$). Second, the crystal-field splitting is responsible for an orientation dependent transition frequency in that regime, depending on the angle between the symmetry axes and the external field $\vartheta_i$ and scaling as $\propto D\,\cos2\vartheta_i$. As a consequence, the transition frequencies of individual NV-centers differ by typical values of several $100\,{\rm MHz}$, thus providing the possibility for individual microwave addressing for the values of $\Omega$ obtained from figure\,\ref{b_dynamdecoupl} and at the same time maintaining a uniform dipolar coupling interaction. Moreover these inhomogeneous energy splittings justify the omission of dipolar flip-flop interaction terms in\,(\ref{htot}),  that would not be energy conserving in our setup.

Combining the dipolar interaction with decoupling suppresses the environmental coupling, i.e. decoherence, but also part of the gate interaction, a general problem in the application of decoupling sequences. Transforming to an interaction picture with respect to the driving in the relevant limit  $\Omega\gg J_{ij}$, the effective interaction up to non-nearest neighbours is given by
$H_{I,\mathcal{M}_k}=1/2\,\sum_{(i,j)} J_{ij}\,S_{\mathcal{M}_k}^{ij}$, $S_{\mathcal{M}_1}^{ij}=s_+^i\,s_-^j+\text{h.c.}$ and $S_{\mathcal{M}_2}^{ij}=s_+^i\,s_+^j+\text{h.c.}$,
wherein $(i,j)$ sums over all neighbouring NV-centers and $\mathcal{M}_1$ corresponds to the situation of a homogeneous decoupling field $\Omega_i=\Omega$, whereas $\mathcal{M}_2$ corresponds to the situation $\Omega_i=-\Omega_j$ for $i\in\text{neighb(}j\text{)}$. $s_+$ and $s_-$ are the ladder operators in the $\sigma_x$-eigenbasis.\par
\begin{figure}[htb]
\begin{centering}
\includegraphics[scale=0.3]{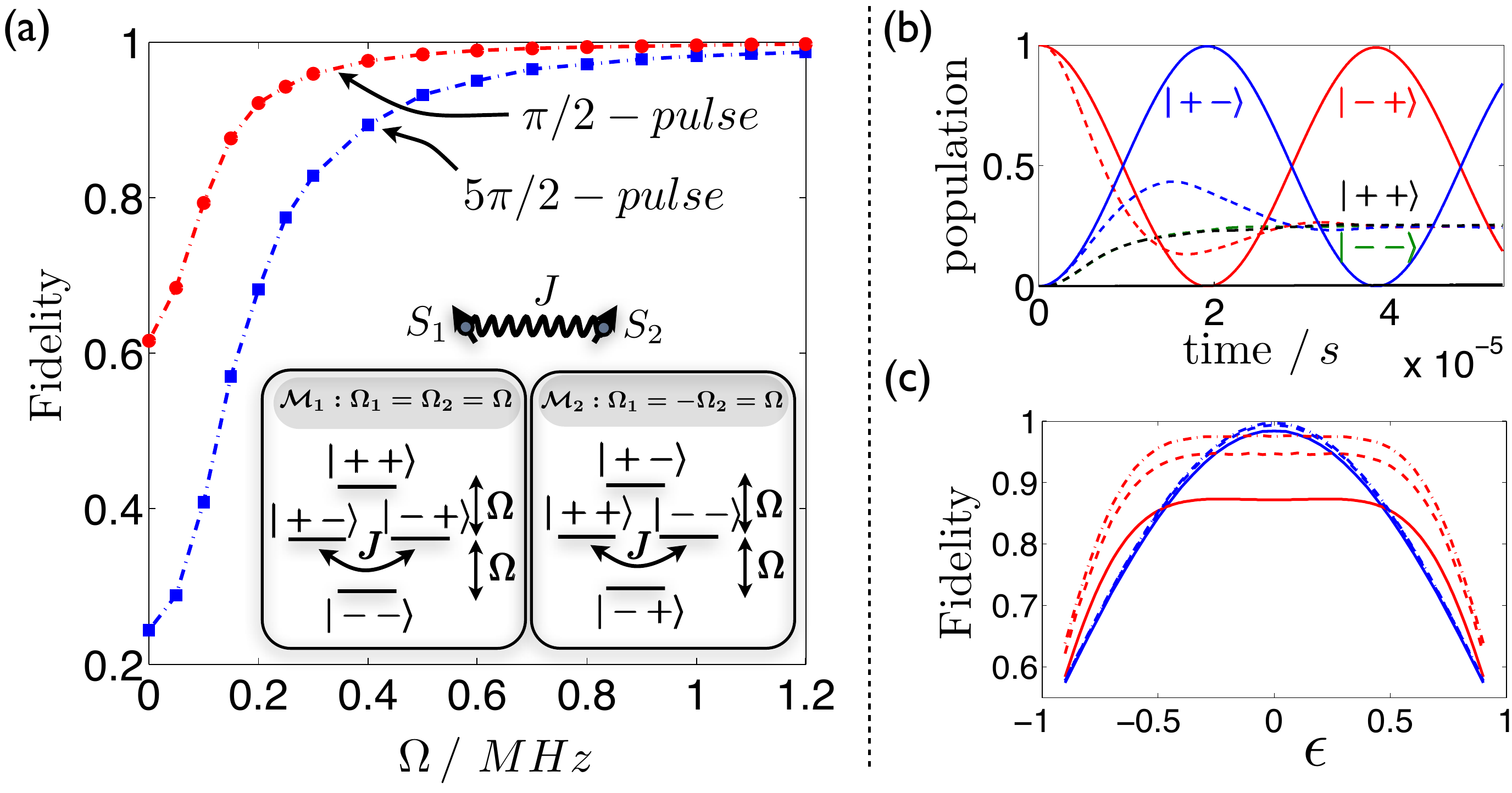}
\caption{\label{b_elspingate} \textbf{Two qubit gates} \textbf{(a)}  Fidelity vs decoupling field strength for a two qubit gate interaction and a $\pi/2$ and $5\pi/2$ rotation in the $\mathcal{M}_1$ manifold leading to a maximally entangled state. Inset: Energy levels and dipolar coupling for the different manifolds. \textbf{(b)} State population vs time for zero decoupling (dashed) and $\Omega=1.2\,{\rm MHz}$ (solid). The former case leads to a maximally mixed state. \textbf{(c)} Fidelity vs systematic error $\epsilon$ for a $\pi/2$ rotation with (red) and without (blue) applying the error compensation sequence for a decoupling field $\Omega=0.5\,{\rm MHz}$ (continuous), $\Omega=0.8\,{\rm MHz}$ (dashed) and $\Omega=1.2\,{\rm MHz}$ (dashed-dotted). The noise parameters are given in figure\,\ref{b_dynamdecoupl} and $J=J_{12}=26\,{\rm kHz}$ corresponding to  the configuration in figure\,\ref{b_dynamdecoupl}\,(c) for $r_{12}=10\,{\rm nm}$.}
\end{centering}
\end{figure}

The two qubit setting is shown in figure\,\ref{b_elspingate}\,(a)-(b), illustrating the fidelity and time evolution to create a maximally entangled state by a $\pi/2$ and $5\pi/2$ two qubit rotation in the $\mathcal{M}_1$\,-\,coupling manifold, respectively, depicted in the left inset of figure\,\ref{b_elspingate}\,(a).  Note that the required time for the latter case exceeds $T_2$($\Omega$=0) by more than a factor of three, therefore impressively demonstrating the decoupling that allows for a 98\% final state fidelity with a decoupling field of $\Omega=1.2\,{\rm MHz}$.
\par
\begin{figure}[htb]
\begin{centering}
\includegraphics[scale=0.26]{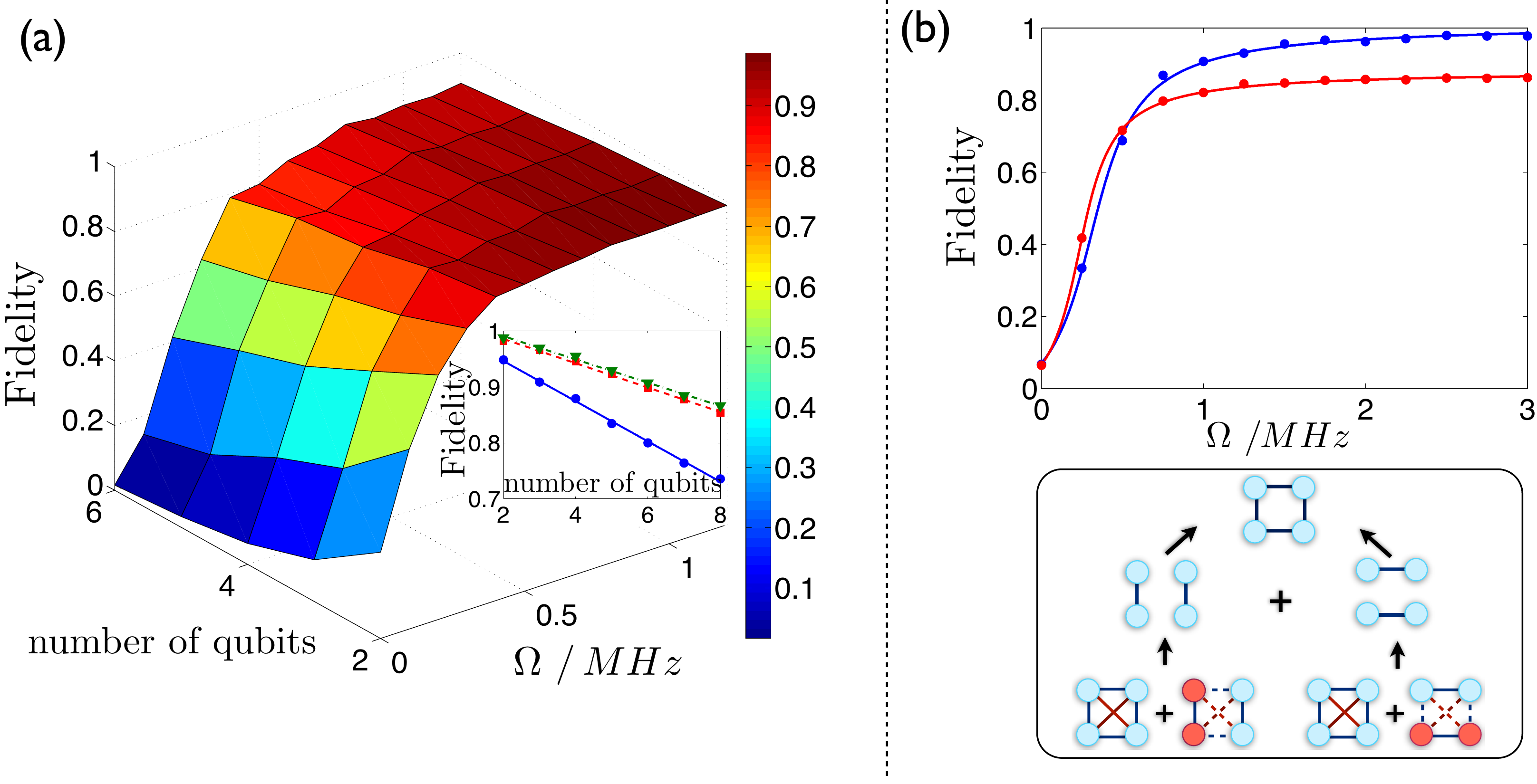}
\caption{\label{b_quantsim} \textbf{Multi qubit gates} (Noise parameters see figure\,\ref{b_dynamdecoupl} and $J_{ij}=J=26\,{\rm kHz}$ corresponding to  the configuration in figure\,\ref{b_dynamdecoupl}\,(c) for $r_{ij}=10\,{\rm nm}$) \textbf{(a)} Fidelity for creating a one dimensional cluster state vs decoupling field strength and the number of qubits involved. Time addition is performed by two Suzuki-Trotter cycles. The inset shows the number scaling for a constant decoupling field of $\Omega=0.5\,{\rm MHz}$ (blue continuous), $\Omega=0.8\,{\rm MHz}$ (red dashed) and $\Omega=1.2\,{\rm MHz}$ (green dashed-dotted). \textbf{(b)} Four qubit 2D cluster state for one (red) and two Suzuki-Trotter cycles (blue) eliminating next nearest neighbour interactions using the addition sequence sketched below: Herein blue lines denote $\sigma_z\otimes\sigma_z$ interactions, red lines $\mathcal{M}_1$ interactions and dashed lines denote a negative sign of the corresponding interactions obtained by red circles representing embedding the interaction in between a $U$, $U^\dagger$ pulse with $U=\text{exp}(-i\,\pi/2\,\sigma_x)$ on the red marked qubits. The basic interactions are obtained by adding $\mathcal{M}_1$ and $\mathcal{M}_2$ as described in the text. }
\end{centering}
\end{figure}
In contrast to these simple two qubit examples, recovering the full dipolar interaction form ($\sigma_z\otimes\sigma_z$-type coupling, see\,(\ref{htot})) is crucial for a wide range of applications ranging from cluster state computation to quantum simulations. Once achieved this allows to create all other types of interactions by merely applying local unitary transformations. The $\sigma_z^i\otimes\sigma_z^j$-type interaction cannot be recovered by any local operation out of the reduced manifold interactions\,$H_{I,\mathcal{M}_k}$; however one can add up the contributions $H_{I,\mathcal{M}_1}$ and $H_{I,\mathcal{M}_2}$ in time by using the Trotter or Suzuki-Trotter formalism to implement the total Hamiltonian (see Supplement, section\,6):
\begin{equation}\label{ising1} H_{zz}=H_{I,\mathcal{M}_1}+H_{I,\mathcal{M}_2}=\sum_{(i,j)}(J_{i,j}/2)\,\sigma_z^i\sigma_z^j\, .\end{equation}
 For this scheme to work, it is crucial that the time addition is based on the interaction frame Hamiltonians $H_{I,\mathcal{M}_k}$ instead of the one defined in\,(\ref{htot}), because only in that case the different timescales of the decoupling and coupling strength allow to create a high fidelity $\sigma_z^i\otimes\sigma_z^j$ gate at the same time preserving the decoherence decoupling effect (see Supplement section\,6.1 for more details). \par
\textbf{Compensation of systematic errors} -- Distance variations between adjacent vacancy centers and different orientations of the symmetry axes make it hard in practice to guarantee a uniform coupling. Therefore the coupling coefficient $J_{ij}$ appearing in $H_{I,\mathcal{M}_k}$ (for the two qubit situation) and $H_{zz}$ may be replaced by $J (1+\epsilon_{ij})$ with $\epsilon_{ij}$ describing the systematic error from the optimal case. However, extending the concepts provided in\,\citep{brown04} allows to construct compensation sequences (see Supplement, section\,7) provided that the error $\epsilon_{ij}\lesssim0.5$ and that non-nearest neighbour couplings can be efficiently suppressed, as can for example be achieved by approaches like the one discussed in figure\,\ref{b_quantsim}\,(b).
We analyzed the compensation method for the two qubit case in figure\,\ref{b_elspingate}\,(c) and applied it to a four qubit cluster state in figure\,S9 (Supplement). Due to the significantly increased time, that is eight and sixteen times the original gate operation, respectively, the region of benefit increases with the decoupling field strength provided that it exceeds a threshold magnitude. As expected the two qubit gate compensation is more efficient providing good results already for a $1.2\,{\rm MHz}$ decoupling field in contrast to the multiqubit counterpart that relies on a more general sequence less efficient in time.\par
\textbf{Cluster-state creation and quantum simulation} -- An interesting application of the concepts developed in the preceding sections is the creation of cluster states. These highly entangled states, defined as the unique eigenstates of the multi-body generators $K^{(i)}=\sigma_x^{i}\otimes_{(j,i)} \sigma_z^{j}$ via the eigenvalue equation $K^{(i)}\,\ket{\phi_\mathcal{C}}=\ket{\phi_\mathcal{C}}\,\forall i$, allow to perform any quantum computation operation by purely local measurements on individual qubits\,\citep{raussendorf03}.  It has been shown\,\citep{raussendorf03} that the product of two-body phase gates $S$ applied to a specific initial product state allows for the creation of such an eigenstate, namely $\ket{\phi}_\mathcal{C}= S\ket{++\dots +}$. The connection to the Ising Hamiltonian\,(\ref{ising1}), that can be realized in the nanodiamond system proposed, follows by noting that $S\simeq \text{exp}\left(-i\,\pi/4\,\sum_{(i,j)}\sigma_z^i\,\sigma_z^j \right)$ up to local operations, the latter being implementable by fast pulses on the electron spin manifold.
 As can be seen in figure\,\ref{b_quantsim} (a) and (b)  the combination of decoupling and time addition is capable of creating one and two dimensional cluster states with fidelities well above 90\%. As a second application making use of the available interactions, the simulation of a Heisenberg-chain Hamiltonian is illustrated in the Supplement figure\,S7, an interesting model for the spin dynamics and magnetism in solid state systems.\par

\textbf{Summary} --- In summary we proposed a new method to create scalable arrangements of NV-centers in diamond by exploiting the ability of biological systems for self-assembly along with the precise positioning of surface functionalized nanodiamonds in such structures. We experimentally realized and verified the creation of ordered nanodiamond structures on a protein scaffold, namely on a SP1 monolayer, as well as the SP1-assisted formation of nanodiamond clusters in solution. Based on the achievable NV distances on the nanometer scale we proposed and analyzed theoretically the implementation of single and multiqubit gates and demonstrated its application for the creation of cluster states, thereby addressing the typical problems as the limited coherence time and heterogeneous dipolar coupling strengths. Moderate decoupling fields around 1MHz, well within reach of current experimental setups, allow for the efficient decoupling from surface spin noise with coherence times comparable to the ultimate $T_1$ limit.  Along with significant dipolar  couplings of several tens of kHz and the viability of individual addressing, gate fidelities well above 95\% can be expected even for multiple qubits and imperfect couplings. We believe that the combination of nanodiamonds with biological systems provides a promising approach towards scalability, overcoming the limitations of current attempts and offering a high level of control in the structure formation.
\clearpage
\textbf{Acknowledgements: } This work was supported by the Alexander von Humboldt Foundation, the BMBF project QuOReP (FK 01BQ1012), the EU STREP project DIAMANT, the ERC Synergy grant BioQ, the GIF project "Non-linear dynamics in ultra-cold trapped ion crystals", the DARPA (QuASAR project), the DFG with the programs SPP 1601/1 "New frontiers in sensitivity for EPR spectroscopy: from biological cells to nano-materials" and CU 44/3-2 "Single Molecule based Ultra High Density Memory" and the Taiwan Academia Sinica Research Program on Nanoscience and Nanotechnology "Bio-inspired Organic/Inorganic Hybrid Nano Electronic devices". We thank the Peter Brojde Center for Innovative Engineering and Computer Science for support as well as the bwGRiD project for computational resources.
\vspace{2ex}\\
\textbf{Author contributions: } MBP proposed the concept and project; AR and MBP coordinated the project. AA performed the theoretical calculations with input from AR and MBP; FJ provided the nanodiamonds; YN \& OS provided the SP1 protein and performed the genetic engineering; GK performed the experiments under the supervision of DP, SY,YP and FJ; All authors discussed the results; AA drafted the paper with guidance and input from AR \& MBP and contributions from all the authors.

\clearpage

\onecolumngrid

\makeatletter
\renewcommand*{\p@section}{}
\renewcommand*{\p@subsection}{}
\renewcommand*{\p@subsubsection}{}
\makeatother

\setcounter{figure}{0}
\setcounter{equation}{0}
\setcounter{page}{1}
\renewcommand{\thefigure}{S\arabic{figure}}
 \renewcommand{\thetable}{S\arabic{table}}
\renewcommand{\bibsection}{\section*{Supplementary References}}
\renewcommand{\thesection}{\arabic{section}}
\renewcommand{\thepage}{S\arabic{page}}
\setcounter{section}{0}
\renewcommand\thesubsection{\thesection.\arabic{subsection}}
\renewcommand\thesubsubsection{\thesection.\thesubsection.\arabic{subsubsection}}
\renewcommand{\theequation}{\thesection.\arabic{equation}}
\noindent

\begin{center} {\Large \textbf{Self-assembling hybrid diamond-biological quantum devices}} \vspace{1ex}\\ {\large Supplementary Information}\vspace{2ex}\\ Andreas Albrecht, Guy Koplovitz, Alex Retzker, Fedor Jelezko, Shira Yochelis, Danny Porath, Yuval Nevo, Oded Shoseyov, Yossi Paltiel and Martin B Plenio  \end{center}
\vspace{10ex}

\setcounter{equation}{0}
\section{SP1 protein complex and experimental details}
One remarkable aspect of biomolecules is their ability to recognize a wide range of substances with a high degree of specificity. This feature of biomolecules has been extensively explored and a variety of artificial peptide aptamers, that specifically recognize various inorganic materials, have been created by selecting binders from random arrays of amino acids displayed on phages or bacteria (combinatorial biological method)\,\citep{Ssarikaya03,Skase04,Snaik02,Snaik02b,Sso09}.

SP1 is a thermally stable protein, originally isolated from poplar trees\,\citep{Swang02}, which self-assembles to an 11 nm ring-shape dodecamer (12-mer). The protein is exceptionally stable under extreme conditions, being resistant to proteolysis, high temperatures, organic solvents and high levels of ionic detergent\,\citep{Swang06,Sdgany04}. SP1 multivalency allows the display of 12 binding sites upon one protein complex, resulting in a stable and strong binding agent.

Fusion of Carbon Nano Tube (CNT) specific binding peptides to SP1 N-terminus by genetic engineering resulted in an SP1 ring with 12 CNT binding sites, 6 on each side of the ring. This enabled the creation of SP1 variants which tightly bind to CNTs to form a stable SP1/CNT complex\,\citep{Swolf12}. In this work we used the same SP1 variant to attach and order the nanodiamond structures.  The carbon $sp^2$-hybridization formed on the surface of the nanodiamonds was used to link the nanodiamonds. 

\begin{figure}[htb]
\begin{centering}
\includegraphics[scale=0.45]{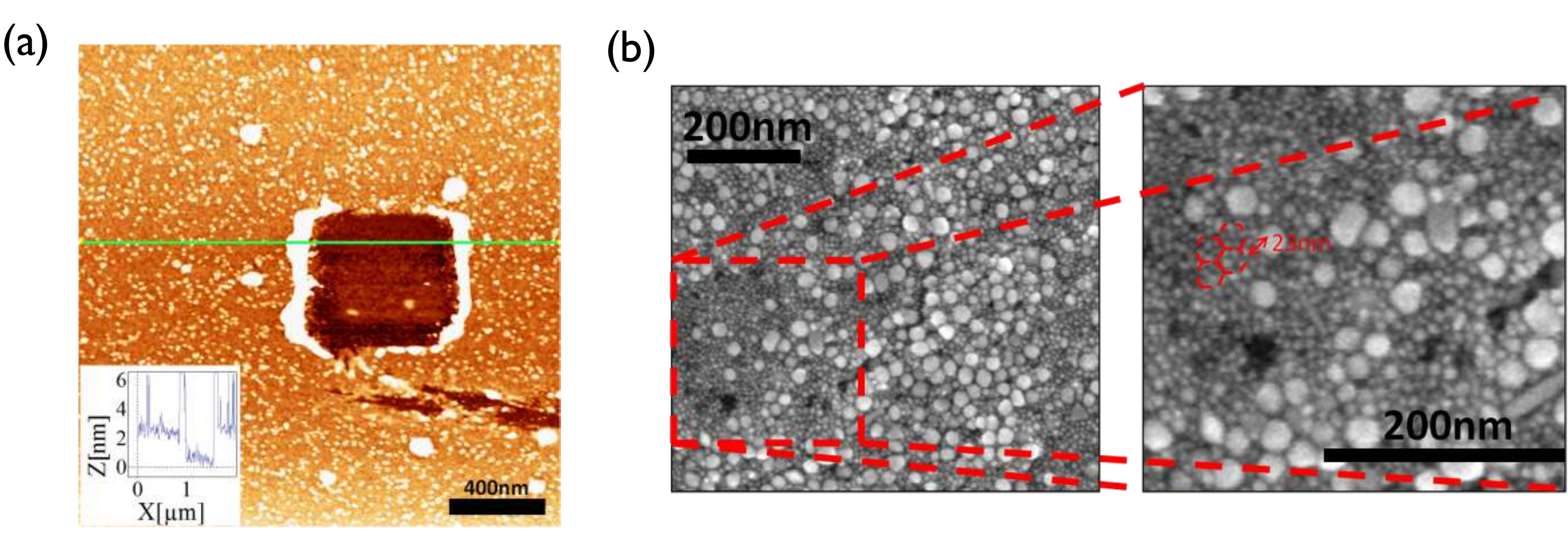}
\caption{\label{gnp_array} \textbf{(a)} AFM scan and line profile (inset) of a dense monolayer of CNT binding SP1 with a scratched area. \textbf{(b)} Small ordered area of 5\,nm gold nanoparticles on a SP1 monolayer formed by the Langmuir-Blodgett method. The periodicity is observed to be 11\,nm, equal to the diameter of the SP1. Alongside the nanoparticles we still have salt particles (the bigger particles), which damage the order and periodicity of the array. The samples were scanned by scanning electron microscopy (Extra High Resolution Scanning Electron Microscopy MagellanTM 400L).}
\end{centering}
\end{figure}

Formation of small ordered nanoparticles areas on an SP1 monolayer is performed using the Langmuir- Blodgett method\,\citep{Sheyman09ii}. In this method a trough is filled with a subphase solution that contains the SP1 proteins and by adding glucose to the subphase solution the SP1 floats to the solution-air interface.
Then, by slowly reducing the interface surface area, the SP1 are forced to compress to a point where they become a monolayer. The SP1 monolayer is transferred to a substrate and washed with distilled water to remove excess salts from the subphase solution. In figure\,\ref{gnp_array}\, (a) an AFM scan of a SP1 layer on a silicon chip with a scratched area is shown. By measuring the height difference between the silicon chip's surface (scratched area) to the rest, we verified the existence of a monolayer of height 2nm, which corresponds to the height of the SP1 protein on a Si surface. This shows that at this stage a dense monolayer of SP1 on the surface has been created. In a second stage the substrate with the SP1 array is inserted into a beaker with nanoparticle solution on an orbital shaker, and afterwards it is washed and dried. In addition to nanodiamonds, experiments with gold particles\,(5nm nanoparticles Sigma Aldrich) have been performed. In that case small ordered areas have been achieved (see figure\,\ref{gnp_array}\,(b)); however excess salts remained and formed salt crystals that lead to lattice defects in the monolayer. As for nanodiamonds (5nm produced by laser ablation from Ray Techniques Ltd) the excess salts were sufficiently removed by cleaning and we observed small periodic hexagonal arrangements of nanodiamonds for dilute particle solutions; however large scale periodic structures have not been achieved yet\,(see figure\,1\,(a)-(b) in the main text). 

\begin{figure}[htb]
\begin{centering}
\includegraphics[scale=0.35]{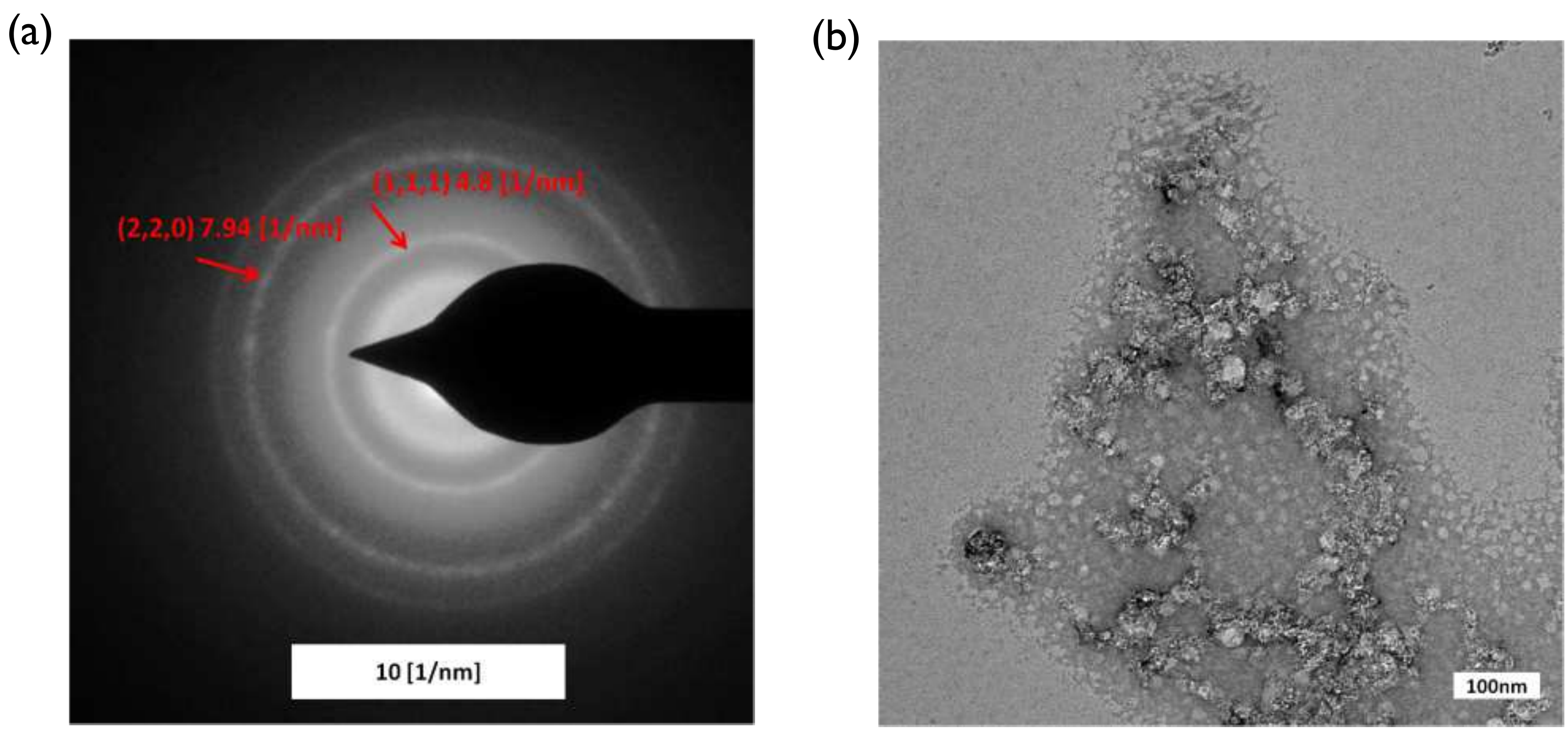}
\caption{\label{em_diff} Electron diffraction image and TEM image of the area where the image was taken.}
\end{centering}
\end{figure}

In order to confirm that the particles (shown in figure\,1 of the main text) are indeed nanodiamonds, electron diffraction measurements have been performed on the sample in the relevant areas. The measurements were taken by the Transmission Electron Microscope Tecnai T12 G$^2$ Spirit. The measured diffraction (1,1,1) lattice parameter results in 2.08\,\AA\, and the (2,2,0) lattice parameter is 1.26\,\AA. This agrees well with the corresponding diamond lattice parameters known from literature and given by 2.04\,\AA\, for (1,1,1) and 1.25\,\AA\, for (2,2,0)\,\citep{Szheng05} (see figure\,\ref{em_diff}). 

The formation of the larger (average nanodiamond size 30\,nm created by grinding) nanodiamond complexes as shown in figure\,1\,(e) of the main text, was achieved by mixing 50\,$\mu$L of 1\,mg/mL of the SP1 solution with 50\,$\mu$L of 1\,mg/mL of the 30\,nm nanodiamonds solution. To the mixture we added 900\,$\mu$L of distilled water. 10\,$\mu$L of the final solution was applied on a silicon chip and scanned by scanning electron microscopy (Extra High Resolution Scanning Electron Microscopy Magellan$^\text{TM}$ 400L).

\clearpage

\setcounter{equation}{0}
\section{Dynamical decoupling and the filter-spectrum overlap approach}\label{s_decoupl}
The effect of dynamical decoupling or more precise the decoherence decay rate of a decoupled system can be described in terms of a filter function overlap, representing the effect of the decoupling field, with the noise spectrum\,\citep{Sgordon07, Salmog11, Sbiercuk2011} (see figure\,\ref{boverlap1}). This allows a very illustrative analysis and description of the working principles of dynamical decoupling methods, opens the possibility of measuring bath-coupling spectra by designing appropriate filter functions and allows for the optimization of decoupling sequences. The only restriction arises from the assumption of the weak coupling limit, i.e. by treating the noise influence in a perturbative way. More practically this means that the coherence time must be large compared to the noise bath correlation time, a condition that -dependent on the type of noise- might not be fulfilled for small decoupling fields. However for the parameters considered in this work, in general $T_2>\tau$, and hence those limitations are not relevant. Moreover the expressions are exact in any limit for the free induction decay and pulsed decoupling schemes provided that the noise is Gaussian as can be easily checked comparing the final expression with the ones derived in a non-perturbative way in e.g.\,\citep{Ssousa09}.  A summary of the, in many cases somewhat lengthy, formulas will be given in section\,\ref{aa_definitions}.
\subsection{Decoherence decay rate}\label{sect_decoh11}
\begin{figure}[htb]
\begin{centering}
\includegraphics[scale=0.5]{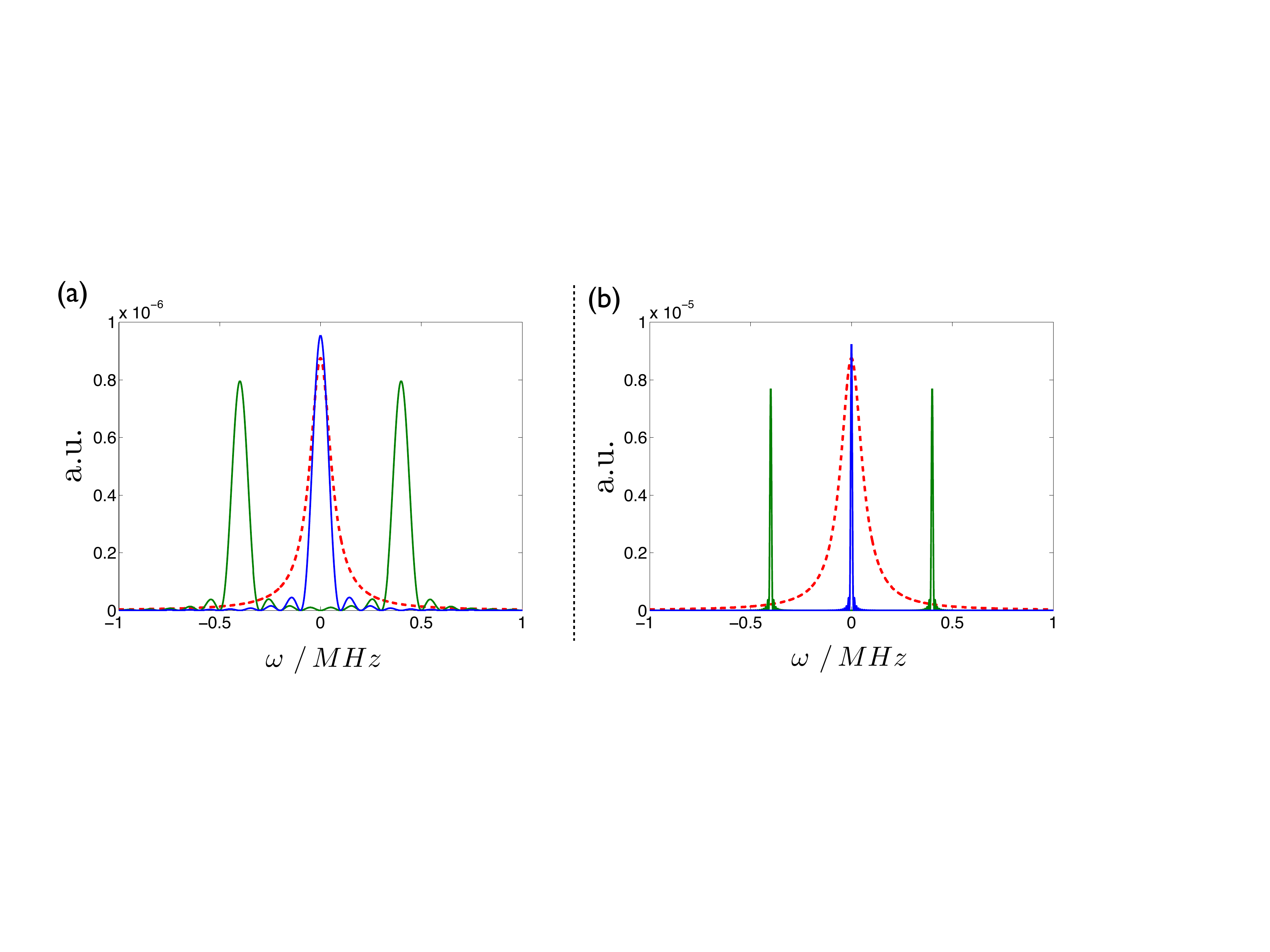}
\caption{\label{boverlap1} Decoupling effect illustrated in the filter - noise spectrum overlap approach for different times (a) $t=10\,\mu s$ and (b) $t=100\,\mu s$. The red dashed curve corresponds to the noise spectrum whereas the blue and green curves represent the filter functions for a zero decoupling field and $\Omega=0.4\,{\rm MHz}$, respectively. For long times $t\gg\tau$ the filter function becomes $\delta$-peaked around $\omega=\pm\Omega$. \textit{Noise parameters:} $T_2=13.3\,\mu s$, $\tau=2.5\,\mu s$, $\sqrt{\ex{\delta^2}}=30.2\,{\rm kHz}$. }
\end{centering}
\end{figure}

\begin{figure}[htb]
\begin{centering}
\includegraphics[scale=0.42]{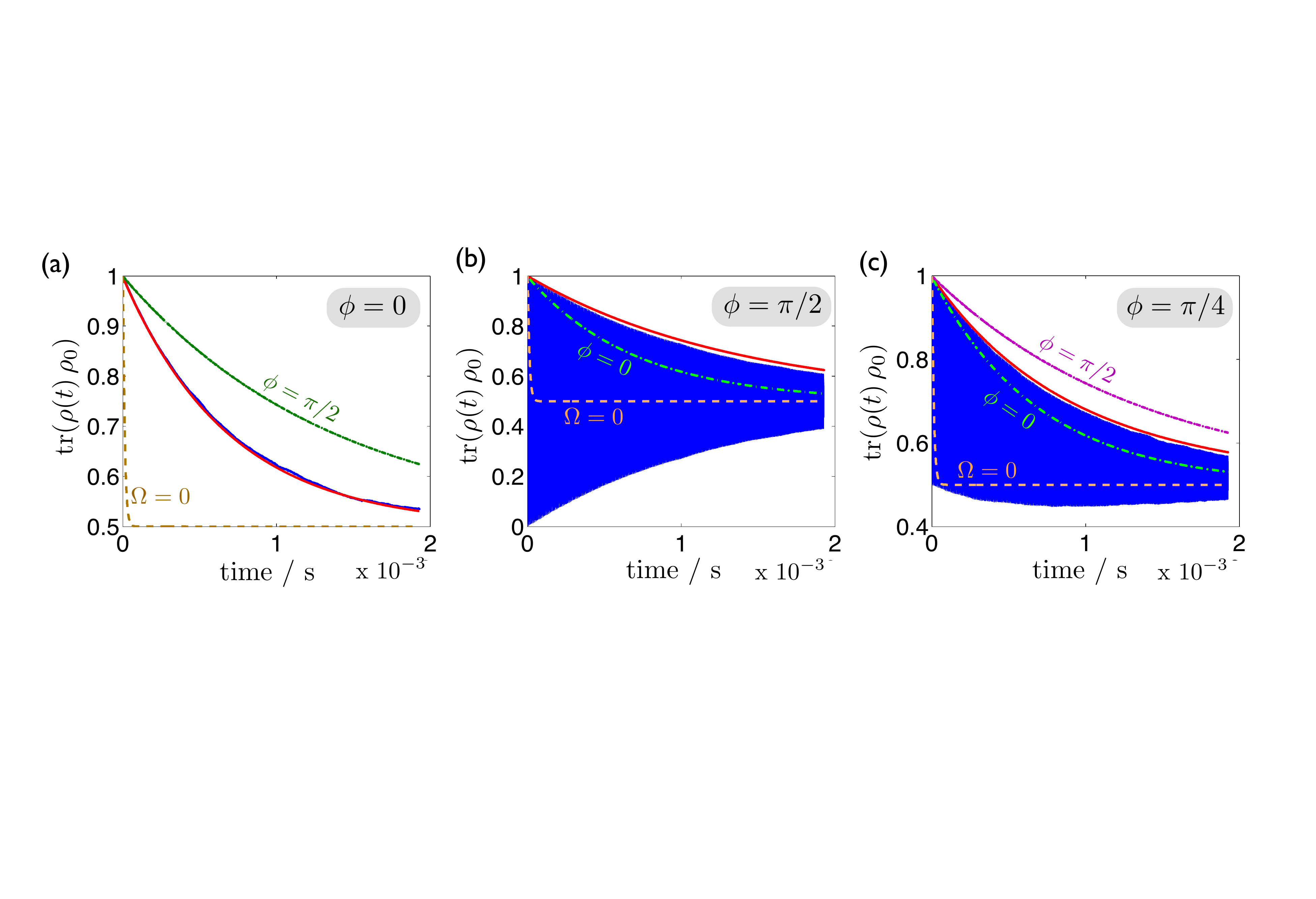}
\caption{\label{bdecohsim} Comparison of the decay rate according to\,(\ref{aa2}) (red lines) to a numerical simulation of the noise process (blue) for an initial state\,(\ref{aa1}) with \textbf{(a)} $\phi=0$, \textbf{(b)} $\phi=\pi/2$ and \textbf{(c)} $\phi=\pi/4$, $\Omega=0.5\,{\rm MHz}$ and modelling the noise as an Ornstein-Uhlenbeck process. Other lines correspond to decay curves for a zero decoupling field and different initial states as indicated in the figure. \textit{Noise parameters:} $T_2=13.3\,\mu s$, $\tau=2.5\,\mu s$, $\sqrt{\ex{\delta^2}}=30.2\,{\rm kHz}$.}
\end{centering}
\end{figure}

Consider a two level system evolving under the Hamiltonian (in the rotating frame)
\begin{equation}\label{noise_ham1}  H=\hbar\dfrac{\delta(t)}{2}\,\sigma_z+\hbar\,\dfrac{\Omega(t)}{2}\,\sigma_x \end{equation}
with $\delta(t)$ a random, zero-mean fluctuating detuning describing the effect of pure dephasing and originating from environmental coupling and $\Omega(t)$ the classical control decoupling field. 
We will assume that the system is initially ($t_0=0)$ prepared in the state
\begin{equation}\label{aa1}\ket{\psi_\phi}=\dfrac{1}{\sqrt{2}}\,\left( \ket{e}+e^{i\,\phi}\,\ket{g} \right)   \end{equation}
whose special cases $\phi=0$ and $\phi=\pi/2$ correspond to the $\sigma_x$ ($\ket{+}_x$) and $\sigma_y$ ($\ket{+}_y$) eigenstate, respectively. Then, after a time $t$ the probability for still finding the system in that initial state is given by
\begin{equation}\label{aa2} \bra{\psi_\phi}\rho\ket{\psi_\phi}=\frac{1}{2}\,\left( 1+\cos^2\phi\,e^{-R_x(t)\,t}+\sin^2\phi\,e^{-1/2\,\left( R_x(t)+R_{\gamma}(t) \right)\,t} \right) \,, \end{equation}
describing purely the effect of decoherence and not taking the coherent evolution into account (see figure\,\ref{bdecohsim}) (It is important to note that the dephasing term contributes as well to the coherent evolution. For example, in the limit $\Omega^2\gg\ex{\delta^2}$ the system performs coherent oscillations with a Rabi frequency $\Omega_{eff}\simeq\omega+\ex{\delta^2}/(2\Omega)$.).\\
The decay rates appearing in\,(\ref{aa2}) can be expressed as
\begin{equation}  R_k(t)=\dfrac{1}{2t}\dfrac{1}{2\pi}\,\int_{-\infty}^\infty\,S(\omega)\,F_t^k(\omega)\,\mathrm{d}\omega \end{equation}
with $S(\omega)=\int_{-\infty}^\infty \text{exp}(-i\omega\tau)\ex{\delta(t)\delta(t+\tau)}\mathrm{d}\tau$ the noise spectrum and $F_t^k(\omega)$ a decoupling field dependent filter function (see figure\,\ref{boverlap1}). Exact expressions can be found in section\,\ref{aa_definitions}.\\
$R_x(t)$ corresponds to the decay rate for an initial $\sigma_x$ eigenstate with $R_x(t)\simeq 1/2\,S(\Omega)$ for $t\gg \tau$ and a constant decoupling field, wherein $S(\Omega)$ is the noise spectrum evaluated at the frequency of the decoupling field. The decay rate $R_y(t)=1/2\,\left( R_x(t)+R_{\gamma}(t) \right)$ corresponds to the decay rate for an initial $\sigma_y$-eigenstate and is split into two parts. 
Splitting the decay rate $R_y(t)$ into the two contributions $R_x(t)$ and $R_{\gamma}(t)$ allows for a more simple interpretation of the decay behaviour. In the limit of $t\gg \tau$ and $\Omega\,t\gg 1$ the contribution $R_{\gamma}(t)\simeq 0$ (more precise $R_{\gamma}(t)\,t\ll R_x(t)\,t$) and thus the decay rate of the second decay contribution corresponds to half the decay rate of the first contribution or as a specific example the decay rate of the $\sigma_x$ is twice as large as the decay rate of the $\sigma_y$ eigenstates. Therefore in that limit ($t\gg\tau$, $\Omega\,t> 1$) and for a constant decoupling field (\ref{aa2}) takes the form:
\begin{equation}\label{dec_decay2}   \bra{\psi_\phi}\rho\ket{\psi_\phi}=\frac{1}{2}\,\left( 1+\cos^2\phi\,e^{-R_x\,t}+\sin^2\phi\,e^{-1/2\, R_x\,t} \right) \quad \text{with } R_x=\dfrac{1}{2}\,S(\Omega) .  \end{equation}
This has a clear interpretation in that the first part (the $\sigma_x$ eigenstate decay) is related to a population decay whereas the second part (the $\sigma_y$ eigenstate decay) corresponds to a decay of the corresponding coherences, a process that happens in the Markovian limit $t\gg\tau$ -also characterized by a time independent decay rate- with half of the population decay rate. In the non-Markovian limit the decay rate of the coherence contributions is increased by an additional factor $R_{\gamma}(t)$. Note also that for the case of free induction decay, i.e. $\Omega=0$, $R_{\gamma}(t)=R_x(t)$ and thus both decay contributions are equal as expected by the isotropy of the system. \\

\subsection{Decay rate derivation}
Following the description in\,\citep{Sgordon07, Salmog11} we will derive the general decoherence decay formula (\ref{aa2}).
It is advantageous to evaluate the decoherence behaviour in the $\sigma_x$ eigenbasis $\ket{\pm}_x\equiv \ket{\pm}$ in which the effect of the decoupling field can be interpreted as creating an energy gap suppressing flipping processes by the (off-resonant) detuning fluctuations. Using the Nakajima-Zwanzig projection operator approach\,\citep{Sbreuer01} allows to obtain a master equation for the system interacting with a noise bath up to second order in the coupling constant for the evolution under (\ref{noise_ham1})\,\citep{Sgordon07} 
\begin{equation} \dfrac{\mathrm{d}\rho}{\mathrm{d}t}=-\dfrac{1}{\hbar^2}\,\int_0^t\mathrm{d}t'\,\phi(t-t')\,\left[S(t),S(t')\,\rho(t)  \right]+\text{h.c.}  \end{equation}
with $\phi(t-t')=\ex{\delta(t)\,\delta(t')}$,
 $S(t)=\hbar/2\,\tilde{\sigma}_z$ and \\$\tilde{\sigma}_z=\text{exp}\left( i\int_0^t \Omega(t')/2\,\mathrm{d}t'\,\sigma_x\right)\,\sigma_z\,\text{exp}\left( -i\int_0^t \Omega(t')/2\,\mathrm{d}t'\,\sigma_x\right)$ the $\sigma_z$ operator in the interaction picture with respect to the decoupling field.
It turns out that coherences and populations are decoupled in the differential equation expressed in the $\sigma_x$-basis states leading to ($\rho_{ij}=\ex{i|\rho|j}$),
\begin{equation}  \dfrac{\mathrm{d}}{\mathrm{d}t}\,\begin{pmatrix} \rho_{++} \\ \rho_{--} \end{pmatrix}  =-\dfrac{1}{2}\,\gamma_1(t)\begin{pmatrix} 1 & -1 \\ -1 &1     \end{pmatrix}\,\begin{pmatrix} \rho_{++} \\ \rho_{--} \end{pmatrix} \end{equation}
wherein (see definitions in section\,\ref{aa_definitions}, $U=\text{exp}(i\int_0^t\Omega(\tau)\,\mathrm{d}\tau)$, $\mathcal{R}$ denotes the real part)
\begin{equation} \gamma_1(t)=\int _0^t\,\mathrm{d}t'\,\phi(t-t')\,\mathcal{R}\left[ U(t)\,U^\dagger(t') \right], \quad R_x(t)=\dfrac{1}{t}\,\int_0^t\mathrm{d}t'\,\gamma_1(t')\,\mathrm{d}t'  \end{equation}
leading to the solutions
\begin{equation}\label{aa7} \begin{split} \rho_{++}(t)&=\dfrac{1}{2}\,\left[  (2\,\rho_{++}^0-1)\,e^{-R_x(t)\,t}+1 \right]  \\
\rho_{--}(t)&=\dfrac{1}{2}\,\left[  (2\,\rho_{--}^0-1)\,e^{-R_x(t)\,t}+1 \right] \,.
  \end{split}\end{equation}
On the other hand the differential equation for the coherences takes the form
\begin{equation} \label{aa8} \dfrac{\mathrm{d}}{\mathrm{d}t}\begin{pmatrix} \rho_{-+}-\rho_{+-}\\\rho_{-+}+\rho_{+-}  \end{pmatrix}=-\dfrac{1}{2}\,\begin{pmatrix} \gamma_1(t)+\gamma_2(t)  & i\,(\mu_1(t)-\mu_2(t)) \\ i\,(\mu_1(t)+\mu_2(t))   & \gamma_1(t)-\gamma_2(t)  \end{pmatrix} \, \begin{pmatrix} \rho_{-+}-\rho_{+-}\\\rho_{-+}+\rho_{+-}  \end{pmatrix}\end{equation}
with the additional definitions (herein $\mathcal{R}$ denotes the real and $\mathcal{I}$ the imaginary part)
\begin{equation} \begin{split}   \gamma_2(t)&=\int _0^t\,\mathrm{d}t'\,\phi(t-t')\,\mathcal{R}\left[ U(t)\,U(t') \right],\quad R_{\gamma}(t)=\dfrac{1}{t}\,\int_0^t\mathrm{d}t'\,\gamma_2(t')\,\mathrm{d}t' \\
\mu_1(t)&=\int _0^t\,\mathrm{d}t'\,\phi(t-t')\,\mathcal{I}\left[ U(t)\,U^\dagger(t') \right] \\
\mu_2(t)&=\int _0^t\,\mathrm{d}t'\,\phi(t-t')\,\mathcal{I}\left[ U(t)\,U(t') \right]\,.
 \end{split}\end{equation}
As will be seen later, the combination $\rho_{-+}-\rho_{+-}$ is responsible for the decay of coherences in the density matrix description. Moreover it is straightforward to show that the off-diagonal elements (the $\mu$-terms) are related to a coherent evolution whereas the diagonal elements describe the decay terms of the corresponding quantities. Since we are not interested in the coherent evolution it is possible to set those off-diagonal contributions to zero resulting in a description of the envelope decay of the quantities.  Therefore one ends up with
\begin{equation}  \dfrac{\mathrm{d}}{\mathrm{d}t}(\rho_{-+}-\rho_{+-})=-\dfrac{1}{2}\,\left[\gamma_1(t)+\gamma_2(t)  \right] \, (\rho_{-+}-\rho_{+-})\end{equation}
\begin{equation}\label{aa11} \begin{split} (\rho_{-+}-\rho_{+-})(t)&=\text{exp}\left( -\frac{1}{2}\,\int_0^t\mathrm{d}t'\left[\gamma_1(t')+\gamma_2(t')  \right] \right)\,(\rho_{-+}-\rho_{+-})(0)\\
&=e^{-1/2\,(R_x(t)+R_{\gamma}(t))\,t}\,(\rho_{-+}-\rho_{+-})(0)\,.
 \end{split}\end{equation}
Now consider the arbitrary phase state\,(\ref{aa1}) that can be expressed in the $\ket{\pm}$ basis as (neglecting a global phase factor)
\begin{equation} \ket{\psi_\phi}=\cos(\phi/2)\, \ket{+}-i\,\sin(\phi/2)\ket{-}  \end{equation}
leading to the initial density matrix
\begin{equation}\label{aa13} \begin{split} \rho_0=\ket{\psi_\phi}\bra{\psi_\phi}&=\ket{+}\bra{+}\,\cos^2(\phi/2)+\ket{-}\bra{-}\,\sin^2(\phi/2)\\
&-i\,\left( \ket{-}\bra{+}-\ket{+}\bra{-}\right)\,\cos(\phi/2)\,\sin(\phi/2).
 \end{split} \end{equation}
Using eqn.\,(\ref{aa7}) and \,(\ref{aa11}), the density matrix after a time evolution of $t$ follows to be
\begin{equation}\label{aa14} \begin{split} \rho(t)&=\ket{+}\bra{+}\,\dfrac{1}{2}\,\,\left[  (2\,\cos^2(\phi/2)-1)\,e^{-R_x(t)\,t}+1 \right] \\
     &+ \ket{-}\bra{-}\,\dfrac{1}{2}\,\left[(2\,\sin^2(\phi/2)-1)\,e^{-R_x(t)\,t} +1  \right]\\
     &-i\,\left( \ket{-}\bra{+}-\ket{+}\bra{-}\right)\,\cos(\phi/2)\,\sin(\phi/2)\,\,e^{-1/2\,(R_x(t)+R_{\gamma}(t))\,t}\,.
 \end{split} \end{equation}
Knowing the density matrix time evolution it is straightforward to calculate the decay rate of the initial state $\ket{\psi_\phi}$ towards the completely mixed state $\rho_{\rm mixed}=1/2\,(\ket{+}\bra{+}+\ket{-}\bra{-})$ and the probability for finding the system in the initial state after a time $t$:
\begin{equation}\label{aa15} \text{tr}\left( \ket{\psi_\phi}\bra{\psi_\phi}\,\rho(t) \right) = \bra{\psi_\phi}\rho\ket{\psi_\phi}=\frac{1}{2}\,\left( 1+\cos^2\phi\,e^{-R_x(t)\,t}+\sin^2\phi\,e^{-1/2\,\left( R_x(t)+R_{\gamma}(t) \right)\,t} \right)\end{equation}
 what corresponds exactly to the result given in eqn.\,(\ref{aa2}).

\subsection{Population decay for $\phi=0$ and $\phi=\pi/2$ and a constant decoupling field}
In this section the limiting cases of starting in the $\sigma_x$ eigenstate $\ket{+}_x$ ($\phi=0$) and the $\sigma_y$ eigenstate $\ket{+}_y$ ($\phi=\pi/2$), respectively, shall be reviewed for the case of a constant decoupling field $\Omega(t)=\Omega=const.$, leading to an explanation of the different decay rates in both basis states.\par
For $\phi=0$ the initial state is given by\,(\ref{aa13})
\begin{equation} \rho_0^x=\ket{+}\bra{+}  \end{equation} 
and the time evolution follows from\,(\ref{aa14}) 
\begin{equation} \rho^x(t)=\ket{+}\bra{+} \,\,\dfrac{1}{2}\,\left( 1+e^{-R_x(t)\,t} \right)+\ket{-}\bra{-}\,\,\,\dfrac{1}{2}\,\left(1-e^{-R_x(t)\,t} \right)   \end{equation}
leading to a decay of the initial state\,(\ref{aa15})
\begin{equation} \bra{+}\rho^x(t)\ket{+}=\dfrac{1}{2}\,\left( 1+e^{-R_x(t)\,t}\right)\,.  \end{equation}
In the $\ket{\pm}$ basis, that is optimal for the interpretation of the decoupling effect, this corresponds to a population decay with an effective rate determined by the flipping term $\delta(t)$ (the pure dephasing in the original state basis) suppressed by the off resonance due to the decoupling field energy splitting $\Omega$.\par
For $\phi=\pi/2$ the initial state takes the form\,(\ref{aa13})
\begin{equation}  \rho_0^y=\dfrac{1}{2}\,(\ket{+}\bra{+}+\ket{-}\bra{-})-\dfrac{i}{2}\,\left(\ket{-}\bra{+}-\ket{+}\bra{-} \right)\,. \end{equation}
Note that the first contribution already corresponds to the completely mixed state and does indeed not change in time such that the decay is determined by the decay of the $\sigma_x$ basis coherences\,(\ref{aa14})
\begin{equation} \rho^y(t)=\dfrac{1}{2}\,(\ket{+}\bra{+}+\ket{-}\bra{-})-\dfrac{i}{2}\,\left(\ket{-}\bra{+}-\ket{+}\bra{-} \right)\, e^{-1/2\,(R_x(t)+R_{\gamma}(t))\,t}\,\end{equation}
leading to the probability for finding the system in the initial state\,(\ref{aa15})
\begin{equation}  \text{tr}\left( \rho_0^y\,\rho^y(t) \right)=\dfrac{1}{2}\,\left( 1+e^{-1/2\,(R_x(t)+R_{\gamma}(t))\,t} \right) \,.\end{equation}
Referring again to the decoupled decay picture it is clear that the decay of the $\sigma_y$ eigenstates is governed by the decay of the $\sigma_x$ coherences. \par
This interpretation  of  pure population ($\phi=0$) and coherence ($\phi=\pi/2$) decay allows especially for a simple explanation of the decay rate difference in the Markovian limit ($t\gg\tau$, $\Omega\,t>1$): In that case the coherence decay induced by the population decay is just half that rate as also follows by noting that in that limit $R_{\gamma}(t)\simeq 0$ and $R_x(t)=1/2\,S(0)=R_x=const$. For a Markovian master equation description that property follows from the conservation of the density matrix trace as can be easily checked analysing the master equation (i.e. the trace of the right hand side should be zero):
\begin{equation} \dfrac{\mathrm{d}\rho}{\mathrm{d}t}= \dfrac{R_x}{4}\,\left( 2\,\sigma_-\rho\sigma_++2\,\sigma_+\rho\sigma_- -\left\{ \sigma_+\sigma_-,\rho  \right\}-\left\{ \sigma_-\sigma_+,\rho  \right\}\right)\,. \end{equation}\par
\clearpage
\subsection{Decay rates, filter functions and definitions}\label{aa_definitions}
\textbf{Decoherence rate}\\
\textit{Definition:} 
\[ U(t):= e^{i\,\int_0^t\Omega(t')\,\mathrm{d}t'} \]
\[  \phi(t-t')=\ex{\delta(t)\,\delta(t')}  \]
\begin{equation*}\begin{split}  R_x(t)&=\dfrac{1}{t}\,\int_0^t\,\mathrm{d}t'\int_0^{t'}\,\mathrm{d}t''\,\phi(t'-t'')\,\mathcal{R}\left(U(t')\,U^\dagger(t'') \right)\\
		&=\dfrac{1}{t}\,\int_0^t\,\mathrm{d}t'\int_0^{t'}\,\mathrm{d}t''\,\phi(t'-t'')\cdot\\
&\quad\qquad\,\cdot\left[ \cos\left(\int_0^{t'}\,\mathrm{d}\tau\,\Omega(\tau)  \right)\,\cos\left(\int_0^{t''}\,\mathrm{d}\tau\,\Omega(\tau)  \right)+\sin\left(\int_0^{t'}\,\mathrm{d}\tau\,\Omega(\tau)  \right)\,\sin\left(\int_0^{t''}\,\mathrm{d}\tau\,\Omega(\tau)  \right)   \right]
 \end{split} \end{equation*}
\begin{equation*}\begin{split}  R_{\gamma}(t)&=\dfrac{1}{t}\,\int_0^t\,\mathrm{d}t'\int_0^{t'}\,\mathrm{d}t''\,\phi(t'-t'')\,\mathcal{R}\left(U(t')\,U(t'') \right)\\&=\dfrac{1}{t}\,\int_0^t\,\mathrm{d}t'\int_0^{t'}\,\mathrm{d}t''\,\phi(t'-t'')\cdot\\
&\quad\qquad\,\cdot\left[ \cos\left(\int_0^{t'}\,\mathrm{d}\tau\,\Omega(\tau)  \right)\,\cos\left(\int_0^{t''}\,\mathrm{d}\tau\,\Omega(\tau)  \right)-\sin\left(\int_0^{t'}\,\mathrm{d}\tau\,\Omega(\tau)  \right)\,\sin\left(\int_0^{t''}\,\mathrm{d}\tau\,\Omega(\tau)  \right)   \right]
 \end{split} \end{equation*}
\textbf{Filter functions}
\[R_i(t)=\dfrac{1}{2\,t}\,\dfrac{1}{2\,\pi}\,\int_{-\infty}^\infty\,S(\omega)\,F_t^i(\omega)\,\mathrm{d}\omega  \]
\textit{with}
\[  S(\omega)=\int_{-\infty}^\infty\,e^{-i\,\omega\,\tau}\,\ex{\delta(t)\,\delta(t+\tau)}\,\mathrm{d}\tau\]
\textit{and}
\[ F_t^x(\omega)= \left| \int_0^t\,e^{-i\,\omega\,t'}\,\cos\left(\int_0^{t'}\,\mathrm{d}\tau\,\Omega(\tau)  \right)\,\mathrm{d}t'   \right|^2+ \left| \int_0^t\,e^{-i\,\omega\,t'}\,\sin\left(\int_0^{t'}\,\mathrm{d}\tau\,\Omega(\tau)  \right)\,\mathrm{d}t'   \right|^2 \]
\[ F_t^\gamma(\omega)= \left| \int_0^t\,e^{-i\,\omega\,t'}\,\cos\left(\int_0^{t'}\,\mathrm{d}\tau\,\Omega(\tau)  \right)\,\mathrm{d}t'   \right|^2- \left| \int_0^t\,e^{-i\,\omega\,t'}\,\sin\left(\int_0^{t'}\,\mathrm{d}\tau\,\Omega(\tau)  \right)\,\mathrm{d}t'   \right|^2 \]
\textbf{Normalized filter functions \quad$\int_{-\infty}^\infty\,F_t^{k,norm}(\omega)\,\mathrm{d}\omega=1$}
\[  F_t^{x,norm}(\omega)=\dfrac{1}{2\,\pi\,t}\,F_t^x(\omega)  \]
\[ F_t^{\gamma,norm}(\omega)=\dfrac{1}{\mathcal{N}_\gamma}\,F_t^\gamma(\omega)\quad \text{with }\mathcal{N}_\gamma=2\,\pi\,\int_0^t\,\mathrm{d}t'\left[ 2\,\cos^2\left( \int_0^{t'} \Omega(\tau)\mathrm{d}\tau \right)  -1\right]  \]

\textbf{Specific filter functions:}
\begin{itemize}
\item\textbf{Free induction decay ($\Omega=0$)}
\[   F_t^x(\omega)=F_t^\gamma(\omega) = \dfrac{4\,\sin^2\left(\frac{\omega}{2}\,t\right)}{\omega^2}\]
\[ F_t^{x,norm}(\omega)=F_t^{\gamma,norm} (\omega)=\dfrac{1}{2\,\pi\,t}\,F_t^x(\omega)=\dfrac{1}{2\,\pi\,t}F_t^\gamma(\omega) \]
\[  \lim_{t\to\infty} F_t^{norm}(\omega)=\delta(\omega),  \qquad  \lim_{t\to\infty} R_x(t)= \lim_{t\to\infty} R_\gamma(t)=\dfrac{1}{2}\,S(0) \]
\item\textbf{Continuous control field ($\Omega=const.$)}

\[  F_t^x(\omega)=\dfrac{1}{2}\,t^2\,\left[  \text{sinc}^2\left(\dfrac{\omega-\Omega}{2}\,t  \right)+ \text{sinc}^2\left(\dfrac{\omega+\Omega}{2}\,t  \right) \right]  \]
\[  F_t^\gamma(\omega)=\dfrac{2\,\cos(\Omega\,t)\,\left( \cos(\Omega\,t)-\cos(\omega\,t) \right)}{\omega^2-\Omega^2} \]
\[   F_t^{x,norm}(\omega)=\dfrac{1}{2\,\pi\,t}\,F_t^x(\omega)\]
\[   F_t^{\gamma,norm}=\dfrac{\Omega}{\pi\,\sin(2\,\Omega\,t)}\,F_t^\gamma(\omega) \]
\textbf{\textit{Markovian limit: $t\gg\tau$, $\Omega\,t>1$}}
\[   \lim_{t\to\infty} F_t^{x,norm}(\omega)=\dfrac{1}{2}\left[\delta(\omega-\Omega)+\delta(\omega+\Omega)\right]\]
\[\lim_{t\to\infty} R_x(t)=\dfrac{1}{2}\,S(\Omega) \] 
\[  \lim_{t\to\infty} R_\gamma(t)=0 \]
\end{itemize}

\textbf{Spectrum for the Ornstein Uhlenbeck process}
\[ \ex{\delta(t)\,\delta(0)}=b^2\,e^{-\frac{|t|}{\tau}}  \]
\[S(\omega)=\dfrac{2\,b^2\,\tau}{1+\omega^2\,\tau^2}  \]

\clearpage
\setcounter{equation}{0}
\section{Hamiltonian, pseudospin 1/2 description and quantization axis adjustment using an external magnetic field}\label{sect_hamderiv}
Let us first consider a system of NV centers, namely the electron spin-1 ground state triplet manifold ($^3A$) coupled by a dipolar interaction:
\begin{equation}\label{hammi1}    H=H_0+H_{\rm dip}  \end{equation}
with the zero-field and external magnetic field contribution\,\citep{Sneumann10}
\begin{equation}\label{h_2}    H_0=\sum_i \vec{S}_i\, \textbf{D}_i\,\vec{S}_i+\gamma_{el}\,\vec{B}\,\vec{S}_i  \,. \end{equation}
Herein $\textbf{D}$ denotes the orientation dependent zero-field splitting tensor, $S_i$ the spin-1 operators, $\vec{B}$ the external magnetic field and $\gamma_{el}$ the gyromagnetic ratio of the NV center electron spin. In the principal axis system defined by the NV symmetry axis, the zero-field splitting tensor $\textbf{D}$ is diagonal and takes the form:
\begin{equation}\label{ham3} D_i=\text{diag}\left(-\frac{1}{3} D+E,-\frac{1}{3} D-E, \frac{2}{3}\,D \right)  \end{equation}
with $D=2.87\,{\rm GHz}$ and $E$ introducing an additional coupling capable of lifting the $\ket{\pm1}$ state degeneracy. In nanodiamonds the strain dependent $E$ contribution can be nonzero in contrast to the bulk diamond situation\,\citep{Stisler09}. In the principal axis frame (denoted by the primed quantities) $H_0$ can be rewritten as
\begin{equation}\label{h0_nv}  H_0=\sum_i D\,\left[ S_z'^2-\frac{1}{3}\vec{S'}^2\, \right]+E\,\left[  S_x'^2-S_y'^2\right]+\gamma_{el}\,\vec{B}'\cdot\vec{S}'_i \end{equation}
providing a good choice for either small magnetic fields or in cases where the magnetic field is parallel to the NV center symmetry axis. An arbitrary choice of the reference frame (and therefore the quantization in that case), e.g. choosing a constant laboratory frame for multiple qubits with different symmetry axes, requires to rotate the tensor correspondingly such that it takes a different form for each of the NV's (unless they do have the same orientation).\\ 
The dipolar coupling term has the form\,\citep{Slevitt08}
\begin{equation}\label{h_5}  H_{\rm dip}=\sum_{i>j}\dfrac{\mu_0}{4\pi}\,\dfrac{\gamma_{el}^2\,\hbar}{r_{ij}^3}\,\left[ \vec{S}_i\,\vec{S}_j-3\,\left(\vec{S}_i\cdot\vec{e}_{ij}  \right)\,\left(\vec{S_j}\cdot\vec{e}_{ij}\right) \right]\,. \end{equation}
with $\mu_0$ the magnetic permeability and $r_{ij}$ and $\vec{e}_{ij}$ the distance and unit direction vector between NV centers $i$ and $j$, respectively. Assuming an equal quantization axis and imposing the secular approximation allows to approximate eqn.\,(\ref{h_5}) by
\begin{equation}\label{h_6}\begin{split} H_{\rm dip}&\simeq \sum_{i> j} \dfrac{1}{2}\,\left( \dfrac{\mu_0}{4\pi}\,\dfrac{\gamma_{el}^2\,\hbar}{r_{ij}^3}\right)\,\left( 1-3\,\cos^2\theta_{ij} \right) \,\left[ 3\,S_i^z\,S_j^z-\vec{S}_i\,\vec{S}_j\, \right]\,\\
	&=\sum_{i>j}\dfrac{1}{2}\,\left( \dfrac{\mu_0}{4\pi}\,\dfrac{\gamma_{el}^2\,\hbar}{r_{ij}^3}\right)\,\left( 1-3\,\cos^2\theta_{ij} \right) \,\left[ 2\,S_i^z\,S_j^z-\left(S_i^x\,S_j^x+S_i^y\,S_j^y\,\right) \right]\,.
  \end{split}\end{equation}
with $\theta_{ij}$ the angle between $\vec{e}_{ij}$ and the quantization axis that is given by the external magnetic field direction for $B\gg D$.
Note that the dipolar coupling in the present situation is rather small, e.g. ($\mu_0\,\gamma_{el}^2\,\hbar/(4\pi\,r_{ij}^3))=2\pi\cdot 52\,{\rm kHz} $ for $r_{ij}=10\,{\rm nm}$ and therefore small inhomogeneous broadening effects (e.g. different strain contributions) as well as different orientations of the NV center symmetry axis as will be outlined below, essentially reduce\,(\ref{h_6}) to
\begin{equation} \label{h_7} H_{\rm dip}\simeq \sum_{i>j}\left( \dfrac{\mu_0}{4\pi}\,\dfrac{\gamma_{el}^2\,\hbar}{r_{ij}^3}\right)\,\left( 1-3\,\cos^2\theta_{ij} \right) \,S_i^z\,S_j^z \,.  \end{equation}\par
Creating an arrangement of nanodiamonds has the disadvantage that the symmetry axis of the NV-center, the axis pointing along `N-V' and forming a natural quantization direction due to its associated crystal-field energy splitting $D$ in that direction, is not controllable, i.e. there exists no common quantization axis in a laboratory frame. This originates from the fact that the symmetry axis of individual NV centers in our setup have an equal probability of pointing in any arbitrary spatial direction.  Without or with a small magnetic field $\gamma_{el}\,B\ll D$, the evaluation of the dipolar coupling \,(\ref{h_5}) in the principal axis symmetry frames of the centers involved,  leads to a coupling that depends as well on the relative orientation of the NV symmetry axes as on the one to the vector connecting the centers, therefore leading to a vast distribution of coupling frequencies inside a multi-qubit array.  To solve that disadvantage an applied external magnetic field ($\gamma_{el}\,B > D$), sufficiently strong in the sense that its associated energy shift ($\gamma_{el}\,B$) outperforms the one of the crystal field $D$, redefines a new and common quantization axis. Considering the limiting case $\gamma_{el}\,B\gg D$ ($B\gtrsim 1T$) with the magnetic field pointing in the z-direction,  allows to rewrite Hamiltonian\,(\ref{hammi1}) as (neglecting off-resonant coupling terms of the zero-field splitting contribution)
\begin{equation}  H\simeq\sum_i \dfrac{1}{4}\,\left( S_z^2-\dfrac{1}{3}\,\vec{S}^2 \right)\,\left[ D\,\left( 1+3\,\cos(2\vartheta_i) \right)+3\,E\,\left( 1-\cos(2\,\vartheta_i) \right)  \right] +\gamma_{el}\,B\,S_z+ H_{\rm dip}\end{equation}
with $\vartheta_i$ the angle between the magnetic field direction $z$ and the symmetry axis of NV center $i$ and $H_{\rm dip}$ given by\,(\ref{h_5})-(\ref{h_7}) with $\theta_{ij}$ the angle of the vector connecting $i$ and $j$ to the z-axis, i.e. the external magnetic field. Two important prerequisites are achieved that way: First the quantization axis is now completely determined by the external magnetic field and not by the symmetry axis any more such that the dipolar coupling will be homogeneous and independent of the individual orientations (for equal $\theta_{ij}$). Second it provides individual addressability as there exists an orientation dependent ($\vartheta_i$) distribution of transition frequencies differing in the $10-100\,{\rm MHz}$ range. Moreover this property of individual addressing is directly linked to the suppression of exchange flip-flop terms in the dipolar coupling, simplifying the dipolar coupling to the form\,(\ref{h_7}). Recalling that the mutual dipolar coupling is of the order of several tens of kHz ($\sim 52\,{\rm kHz}$ for a distance of $10\,{\rm nm}$) the much higher differences in the transition frequencies as stated above will make these (in our case non-energy conserving) transitions very unlikely. Thus in the regime of our proposal we can safely describe the coupling as a pure $\sigma_z\otimes\sigma_z$-coupling similar to\,(\ref{h_7}).\\
In the more general case of intermediate magnetic fields the effect on the dipolar couplings is more complicated and cannot be simply described in the $S_x,S_y,S_z$ basis manifold. We analyzed this case numerically below for the relevant quasi-particle two-level system. Effectively, recalling that there will be a continuous microwave driving to reduce decoherence, only a reduction to two states of the Spin-1 system is relevant, i.e. $S_j^z=\ket{+1}\bra{+1}+1/2\,(\sigma_j^z-\mathds{1}_j)$ or $S_j^z=\ket{-1}\bra{-1}+1/2\,(\sigma_j^z+\mathds{1}_j)$ for a zero magnetic field or a magnetic field parallel to the NV center symmetry axis, and therefore the two level reduction allows to rewrite\,(\ref{h_7}) as 
\begin{equation}   \label{h_9} H_{\rm dip}^{\rm TLS}\simeq\sum_{i>j} \left( \dfrac{\mu_0}{4\pi}\,\dfrac{\gamma_{el}^2\,\hbar}{r_{ij}^3}\right)\,\left( 1-3\,\cos^2\theta_{ij} \right) \, \dfrac{1}{4}\,\left(\sigma_i^z\,\sigma_j^z\mp \sigma_i^z\mp\sigma_j^z \right)\,.    \end{equation}
Note that the single flipping terms can be incorporated in the energy Hamiltonian $H_0$, will be suppressed anyway when adding a sufficiently strong continuous decoupling driving field or can also be removed exactly by an echo sequence as discussed in section\,\ref{sect_szsz}.\par
\begin{figure}[htb]
\begin{centering}
\includegraphics[scale=0.38]{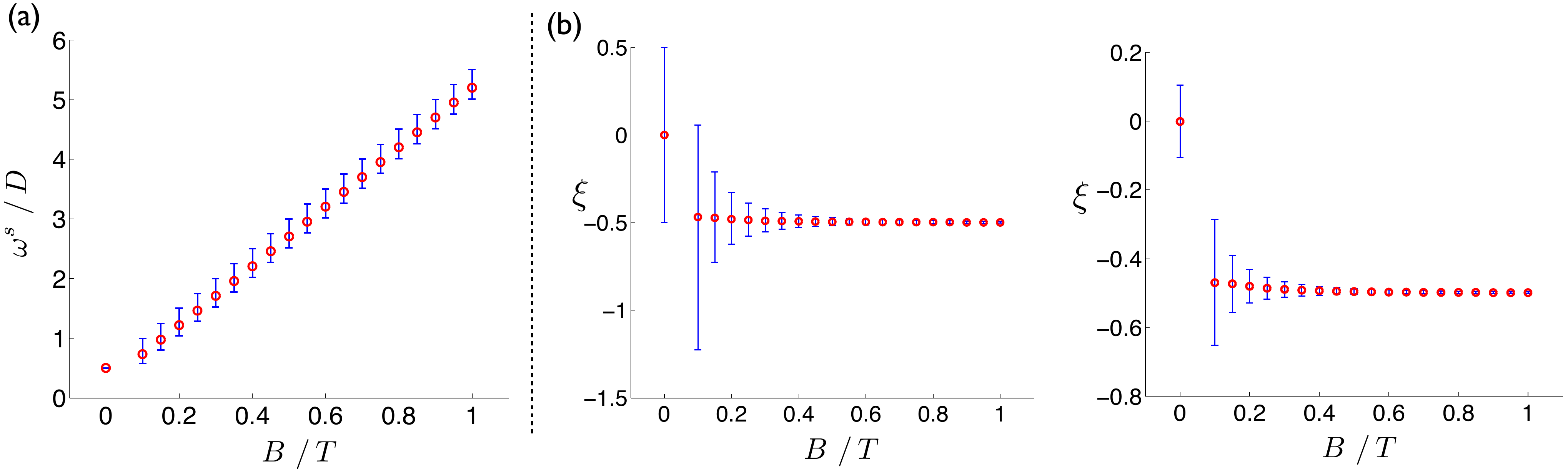}
\caption{\label{bmfieldquanti} \textbf{(a)} Dressed state energy gap $\omega^s$ in units of the zero field splitting $D=2.8\,{\rm GHz}$ for different magnetic field strengths. The two level system was chosen as the transition of neighbouring dressed state energy levels with the largest energy gap. Blue lines indicate the absolute deviations depending on the symmetry axis orientation. \textbf{(b)} Dipolar coupling parameter $\xi$ vs the external magnetic field parallel to the axis connecting the NV centers. The red circles denote the mean values and the blue lines indicate the absolute deviation (left) and variance (right) due to the random orientation of the NV symmetry axis. All values are obtained by averaging over $5\cdot10^5$ random axis orientations. }
\end{centering}
\end{figure}

Adding noise and an additional continuous driving for decoupling, the total Hamiltonian in the two level basis can now be written as
\begin{equation} \label{h_10}   H_{\rm tot}^{\rm TLS}= H_0^{\rm TLS}+H_{n}^{\rm TLS}+H_{\rm drive}^{\rm TLS}+H_{\rm dip}^{\rm TLS}  \end{equation}
with $H_0^{\rm TLS}$ the energy of the two selected dressed states of\,(\ref{h_2}), i.e. obtained by diagonalizing\,(\ref{h_2})
\begin{equation}\label{h_11} H_{0}^{\rm TLS}=\sum_i\dfrac{\omega_i^s}{2}\,\sigma_z^i  \end{equation}
and $\omega_i^s=(D+\gamma_{el}\,B)/2$  for a zero magnetic field or a magnetic field aligned with the NV symmetry axis (and E=0) or $\omega_i^s= (1/8)\,\left[ D\,\left( 1+3\,\cos(2\vartheta_i) \right)+3\,E\,\left( 1-\cos(2\,\vartheta_i) \right)  \right]+\gamma_{el}\,B/2$
 in case of a strong magnetic field $\gamma_{el}B\gg D$. \\
$H_n^{\rm TLS}$ describes the decoherence influence modelled as a pure dephasing noise
\begin{equation}\label{h_deph5} H_n^{\rm TLS}=\sum_i \dfrac{b_i(t)}{2}\,\sigma_i^z  \end{equation}
with $b_i(t)$ a fluctuating frequency as will be discussed in section\,\ref{sect_decoh}.\\
$H_{\rm drive}^{\rm TLS}$ describes the continuous resonant microwave driving for decoupling on the dressed state two level system\,(\ref{h_11}) and is given by
\begin{equation} H_{\rm drive}^{\rm TLS}=\sum_i\Omega_i\,\cos(\omega_i^s\,t)\,\sigma_i^x  \end{equation}
with $\Omega_i$ the Rabi frequency of the driving.\\
The dipolar interaction in the secular approximation can be, noting again that only $\sigma_i^z\,\sigma_j^z$ terms will in general survive,  written as
\begin{equation}\label{h_14} H_{\rm dip}^{\rm TLS}=\sum_{i>j}\dfrac{J_{ij}}{2}\,\sigma_i^z\,\sigma_j^z\quad \text{with  } J_{ij}=2\,\xi_{ij}\, \left( \dfrac{\mu_0}{4\pi}\,\dfrac{\gamma_{el}^2\,\hbar}{r_{ij}^3}\right) \end{equation}
where in the limiting cases of a strong magnetic field, or more general for equal quantization axes of $i$ and $j$ the parameter $\xi_{ij}$ is given by $\xi_{ij}=1/4\,(1-3\cos^2\theta_{ij})$. For general magnetic fields the parameter ranges are plotted in figure\,\ref{bmfieldquanti} and figure\,2\,(c) in the main text, showing that around $B\gtrsim 0.5\,T$ are required for providing an almost uniform coupling; noting that the variance is even lower there is a high probability that a sufficiently uniform dipolar coupling is already obtained for $B\gtrsim 0.2\,T$. \par
Finally, transforming to a rotating frame with respect to the laser frequency and neglecting counter-rotating terms, (\ref{h_10}) then reads
\begin{equation}\label{htot_tls}  H'\simeq \sum_i \dfrac{b_i(t)}{2}\,\sigma_i^z + \sum_i\dfrac{\Omega_i}{2}\,\sigma_i^x+\sum_{i>j}\dfrac{J_{ij}}{2}\,\sigma_i^z\,\sigma_j^z\,. \end{equation}
In the attempt to achieve a uniform dipolar coupling by means of an external magnetic field the orientation of the magnetic field is crucial as can be seen from\,(\ref{h_7}) and (\ref{h_14}). Claiming that the interaction strength should be equal in all spatial directions the maximal dipolar coupling is achieved in a linear chain and in a two dimensional array when the magnetic field is parallel to the chain ($\xi_{ij}=-1/2$) and orthogonal to the plane ($\xi_{ij}=-1/4$), respectively. In a three dimensional array this cannot be achieved, because for the only choice that provides equal interactions in all directions, the space diagonal, $\cos\theta_{ij}=1/\sqrt{3}$ and therefore the interaction strength equals zero (performing an addition in time with non-equal couplings in all spatial directions can however circumvent that problem). As a last remark it should be noted that the combination of a strong magnetic field and the condition $\cos\theta_{ij}=1/\sqrt{3}$ could be used to decouple the system from the dipolar interaction, e.g. after a cluster state is achieved and further local operations and measurements are desired subsequently.  

\subsection{${\rm NV}^{-}$-center hyperfine-structure and its influence on the decoupled gate interaction}
\begin{figure}[htb]
\begin{centering}
\includegraphics[scale=0.38]{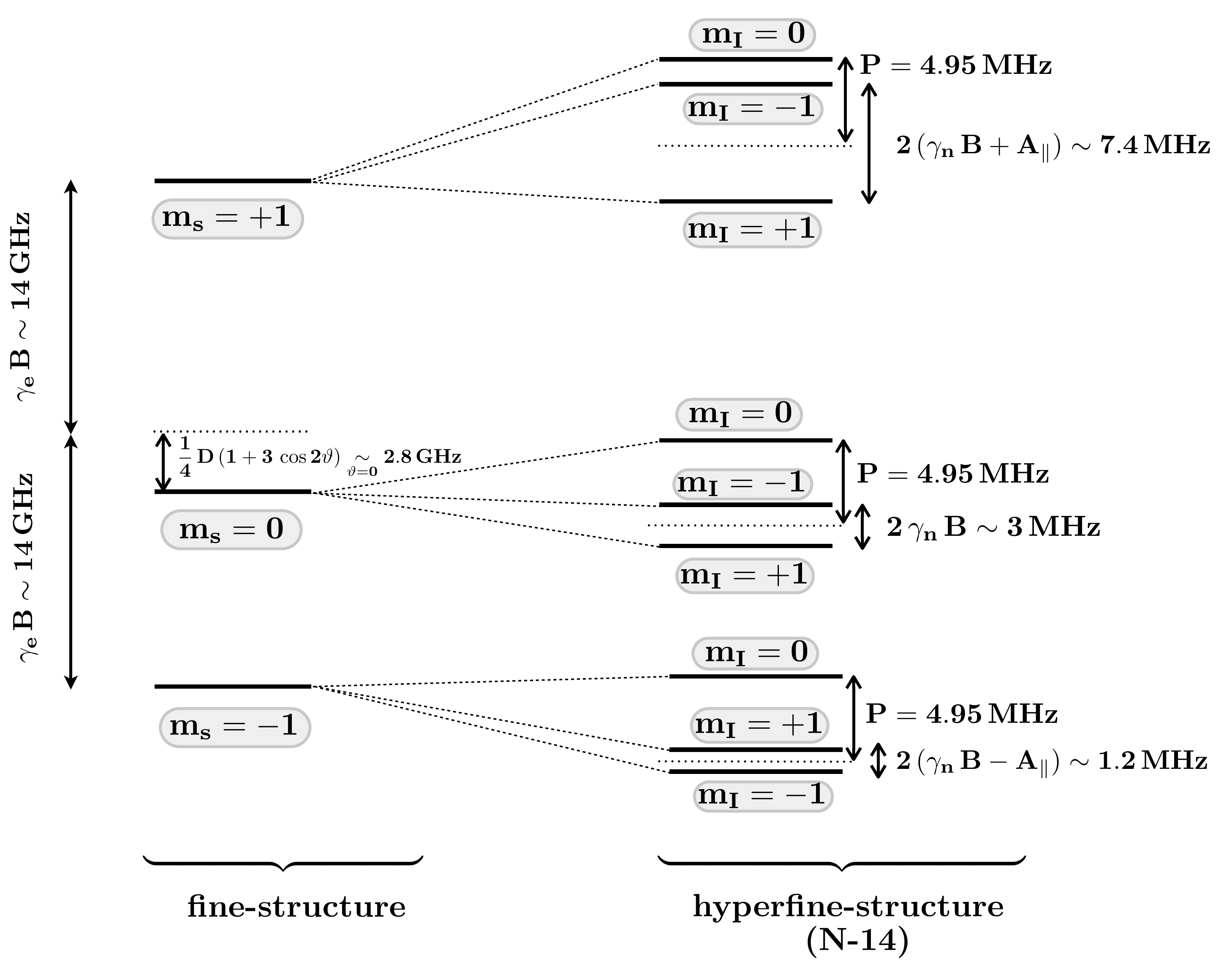}
\caption{\label{b_hyperfine} Fine and hyperfine structure of the ${\rm NV}^-$-center electron spin ground state triplet for an external field $B=0.5\,T$. The nuclear spin consists of the N-14 spin associated with the NV center. The angle $\vartheta$ between the external field and the NV-center symmetry axis is assumed to be sufficiently small in this illustration, allowing to neglect couplings between different nuclear spin states.  }
\end{centering}
\end{figure}

An additional complication, that has been neglected so far, arises from the hyperfine structure associated with the nuclear spin of the nitrogen atom involved in the vacancy center (spin $I=1$ for N-14 or spin $I=1/2$ for the N-15 isotope). In the following we will just consider this `intrinsic' nitrogen nuclear spin in the center itself and assume that other nuclear spins (e.g. $I=1/2$ of C-13) only couple weakly to the center and thus can be treated within the framework of the spin-bath decoherence as outlined in the next section and in the main text. This can be considered as a good approximation for a high purity diamond as well as for strong decoupling fields such that the hyperfine coupling is weak compared to the microwave field, i.e. other nuclear spin are sufficiently far-apart from the NV-center. In that framework the ground-state level structure in the principal axis frame (analogue to (\ref{h0_nv})) can be described as\,\citep{Sacosta11, Ssmeltzer09}
\begin{equation} H_{\rm gs}=H_0+H_{\rm nucl}+H_{\rm hyp}  \end{equation}
with $H_0$ the pure electron spin-Hamiltonian given by\,(\ref{h0_nv}) and $H_{\rm nucl}$, $H_{\rm hyp}$ the (nitrogen) nuclear spin part and the hyperfine coupling interaction, respectively. The nuclear spin Hamiltonian in the presence of a magnetic field $\vec{B}$ takes the form
\begin{equation} H_{\rm nucl}=P\,I_z'^2-\gamma_I\,\vec{B}\vec{I}'   \end{equation}
with the quadrupole splitting $P=-4.95\,{\rm MHz}$ and the gyromagnetic ratio $\gamma_I\simeq 3.1\,{\rm MHz/T}$. For the hyperfine coupling we will neglect exchange (`flip-flop')-terms due to the large energy mismatch between the electron and nuclear spin transitions, such that
\begin{equation}\label{h_hyper}  H_{\rm hyper}\simeq A_\parallel\, S_z'\,I_z'  \end{equation}
with $A=-2.16\,{\rm MHz}$. For the regime of our proposal (B=0.5\,T), this leads to a hyperfine-structure level scheme as depicted in figure\,\ref{b_hyperfine}. With the selection rule $\Delta m_I=0$ it becomes clear that neighbouring transitions between two different electron spin states differ by $\Delta\omega\sim 2.2\,{\rm MHz}$. In order to avoid off-resonant microwave transitions, that would lead to an imperfect decoupling and a different form of the effective dipolar coupling $H_{I,\mathcal{M}_k}$ as is most easily verified in the limit of large detunings, several strategies can be followed: (i) The most simple one consists of eliminating the hyperfine structure by sufficiently large microwave driving fields $\Omega\gg |A_\parallel|$. That way, the hyperfine coupling is suppressed, or analogously all possible nuclear states are excited simultaneously in the strong field limit, and there is thus no need to distinguish between individual hyperfine states. In that limit, the NV-center is described by the electron-spin states to a good approximation and the rather small differences in the hyperfine transitions of $\sim 2{\rm MHz}$ allow to perform such a strong driving by still keeping the microwave power small enough to enable individual addressing (the latter one determined by the rather strong zero-field splitting $\propto 2.8\,{\rm GHz}$).  We would like to point out that continuous microwave drivings of $\sim 40\, {\rm MHz}$ have already been successfully implemented in experiments\,\citep{Scai11}. (ii) A second option would be the regime of weak microwave driving $\Omega\ll |A_\parallel|$. In that regime only a single nuclear spin state transition is excited, provided that the microwave is tuned to resonance with a single specific transition. Off-resonant excitations can be neglected in such a regime and one ends up again with a perfect two-level system. However for the noise parameters considered (see figure\,2 and 3 in the main text), the decoupling in this regime does not lead to optimal results even it might be sufficient for a short $\pi/2$-pulse interaction with tolerable fidelity. As a remark, one might also explore the intermediate regime $\Omega\lesssim |A_\parallel|$ by identifying conditions where the effect of the off-resonant contributions leads to an effective $2\pi$-multiple rotation similar to the technique performed e.g. in\,\citep{Szhao12}. (iii) A third possibility to achieve an effective two-level system consists of polarizing the nuclear spin prior to the actual experiment. Together with the nuclear spin selection rules this results in a single possible transition. Such polarization schemes have been demonstrated in numerous experiments to date as well at room temperature\,\citep{Sjacques09, Ssmeltzer09, Sfischer13} (making use of an excited state level anticrossing that allows direct hyperfine exchange interactions) as at low temperatures using projective measurments\,\citep{Srobledo11} or the concept of coherent population trapping\,\citep{Stogan11}.

\setcounter{equation}{0}
\section{Noise spectrum for a fluorine-terminated nanodiamond surface}
\begin{figure}[htb]
\begin{centering}
\includegraphics[scale=0.38]{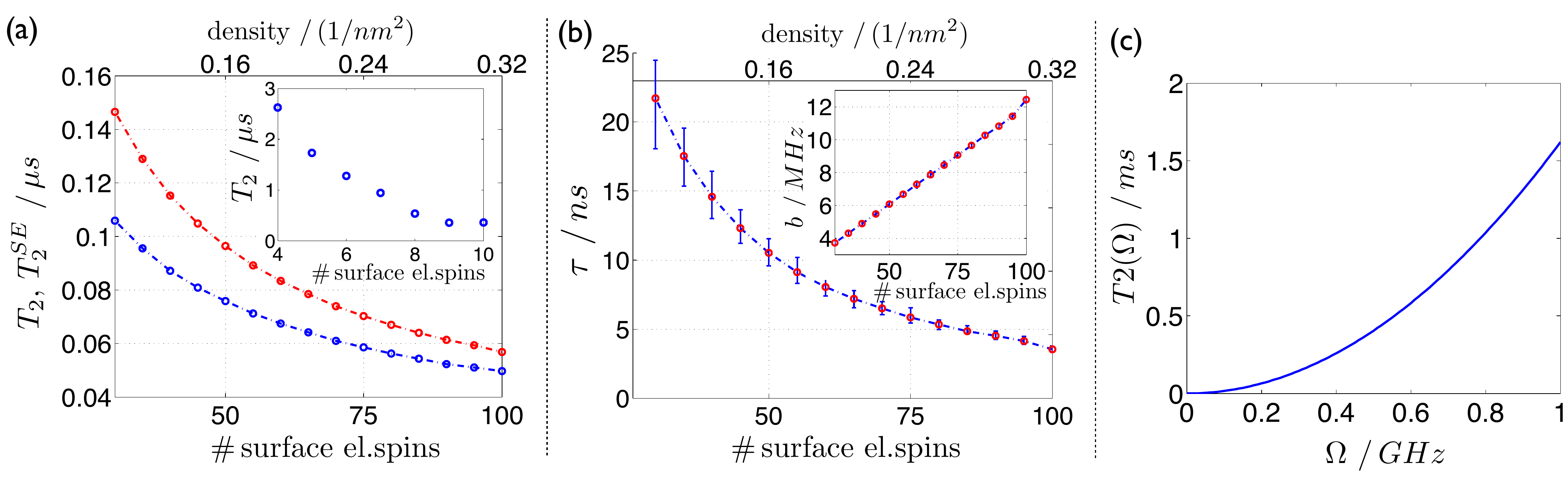}
\caption{\label{belspindecoh} Electron spin decoherence: \textbf{(a)} $T_2$ (blue)  and $T_2^{SE}$ (Hahn-echo, red) time for different numbers of electron surface spins calculated using the mean-field broadening method described in the text. A minimal number of $\sim 30$ spins is required for a Lorentzian-type noise spectrum and the possibility to consider mean values out of the random distribution.  \textbf{Inset:} $T_2$-time out of an exact numerical simulation for a small number of electron spins  emphasizing that $T_2$ times far below $1\mu s$ are expected even for a small number of 10 surface spins. \textbf{(b)} Noise correlation time vs number of surface electron spins. Error bars indicate deviations from the mean value due to the random choice of the electron spin positions. \textbf{Inset:} Noise amplitude vs the number of electron surface spins. \textbf{(c)} Estimated scaling according to\,(\ref{dec_decay2}) of the effective $T_2$ time with the decoupling field $\Omega$ for 30 electron surface spins. The noise parameters are obtained from the mean-field approach and given by $T_2=0.1\,\mu s$, $\tau=21.7\,ns$, $b=3.6\,{\rm MHz}$. For those parameters decoupling fields in the ${\rm GHz}$ range are required which are challenging concerning intensity fluctuation, individual addressing and the validity of the rotating wave approximation. }
\end{centering}
\end{figure}

The main source of decoherence for the nitrogen-vacancy center in diamond arises from couplings of the associated electron spin to the spin bath of additional impurities\,\citep{Szhao12}. For the case of bulk-diamonds, it is well-known that the dominant source of decoherence can be attributed to C-13 nuclear spins (high purity type IIa diamond) and paramagnetic P1-centers associated with nitrogen donors (effective electron spin-1/2 system; type Ib diamond), that naturally occur within the diamond crystal. For nanodiamonds the situation becomes quite different as a result of the surface proximity. In that case surface spins, associated either with the carbon hybridization or the terminating elements, form a dense spin bath that couples with significant strength to the electron-spin of the NV-center\,\citep{Stisler09}. We will therefore focus on this surface induced dephasing mechanism in the upcoming analysis. However it should be noted that other decoherence processes, in particular the mechanisms known already from the bulk diamond counterparts, can and indeed have been described in the same framework\,\citep{Sdobrovitski09, Slange2010} presented in the next two sections. Thus these mechanisms will be decoupled along with the surface-spin decoupling as successfully demonstrated in numerous decoupling experiments to date\,\citep{Slange2010, Snaydenov11, Svandersar12}, and,  keeping in mind that such mechanisms form much smaller spin-bathes, we do not expect changes in the reported $T_2$-times. Here we would also like to point out that even a small number of much stronger coupled electron spins can be decoupled in that framework, as shown at the end of this section and in figure\,\ref{belspindecoh}.\par  
Terminating the nanodiamond surface by fluorine, oxygen or hydrogen / hydroxyl groups replaces the electron spins associated with the sp2 hybridized orbitals by much weaker nuclear spins\,\citep{Shall10} and therefore reduces the dipole-dipole interaction strength between surface spins by a factor of $10^{-6}$ (as given by the square ratio of the corresponding gyromagnetic ratios). Additionally the coupling to the central spin, the electron spin of the NV center, reduces by a factor of $10^{-3}$. Recalling that the pure dephasing noise responsible for the decoherence mechanism can be described by a fluctuating magnetic field and that the strength and fluctuation rate depends on the (hyperfine) coupling to the central spin and the surface spin flip-flop rate, respectively, a significant improvement can be obtained by terminating the surface purely by nuclear spins. As an illustrative example we will concentrate on the fluorine terminated surface (nuclear spin 1/2), having the additional advantage of not affecting the charge state of the NV center and forming a lattice with nearest neighbour distance $2.5\,$\r{A} \,\citep{Sjianming12}. \par
Such a coupled system can be described by
\begin{equation}  H=H_0^{\rm NV}+H_0^{\rm sf}+H_{\rm int}^{\rm sf}+H_{\rm hf} \end{equation}  
with $H_0^{\rm NV}$ the NV center energy Hamiltonian as given in\,(\ref{h_2}), $H_0^{\rm sf}$ the magnetic field splitting of the surface nuclear spins, $H_{\rm int}^{\rm sf}$ the dipole dipole coupling between surface nuclear spins and $H_{\rm hf}$ the hyperfine coupling of the surface spins to the central spin (the NV center electron spin):
\begin{equation}\label{ns_1}\begin{split}  H_0^{\rm sf}&=\sum_i \dfrac{\gamma_{n}\,B}{2}\, \sigma_i^z\\
  H_{\rm int}^{\rm sf}&\simeq\sum_{i>j}\left( \dfrac{\mu_0}{4\,\pi}\,\dfrac{\gamma_{n}^2\,\hbar}{r_{ij}^3}  \right)\,(3\cos^2\theta_{ij}-1)\,\left( -\sigma_i^z\,\sigma_j^z+\frac{1}{2}\,\left[ \sigma_i^x\,\sigma_j^x+\sigma_i^y\sigma_j^y \right] \right)\\
 H_{\rm hf}&\simeq\sum_i \left(\dfrac{\mu_0}{4\,\pi}\,\dfrac{\gamma_{el}\,\gamma_{n}\,\hbar}{r_i^3}  \right)\,(1-3\,\cos^2\theta_i)\,\,\,S^z\,\sigma_z^i \,.  \end{split}\end{equation}
Herein $S$ denotes the spin-1 operator of the vacancy center, $\sigma_k$ the spin 1/2 operators of the surface nuclear spins, $r_{ij}$ the distance between spins $i$ and $j$ and $r_{i}$ the one between surface spin $i$ and the central spin and $\theta_{ij}$ and $\theta_i$ the angles between the vector connecting the two coupled spins involved and the external magnetic field. As discussed in section\,\ref{sect_hamderiv} we will assume that the quantization axis of the NV center is essentially determined by the external magnetic field and not by the symmetry axis of the vacancy center. Alternatively one might assume that the external magnetic field is parallel to the NV symmetry axis. In the two-level approximation, used for the driven system in section\,\ref{sect_hamderiv}, one can substitute $S^z\rightarrow 1/2\,(s_z\pm1)$ with $s_z$ the spin 1/2 operator of the quasi-spin.
Herein we just retained secular contributions of the dipole dipole coupling and neglected flipping terms in the hyperfine interaction (that would be related to $T1$) due to the large energy mismatch between surface spins and the NV electron spins. However, nevertheless those flip flop processes are very unlikely, second order processes might lead to a significant enhancement of the surface spin flips described by $H_{\rm int}^{\rm sf}$ of the order of $A_i^2/\Delta$ with $\Delta$ being the typical energy splitting of the NV levels and $A_i$ the hyperfine coupling amplitude\,\citep{Smaze08,Scywinski09,Syao06}. However we do not expect those terms to be significant in the present proposals that requires magnetic fields $B\gtrsim 0.5\,{\rm T}$  what leads to mediated couplings of the order of $A_i^2/\Omega\sim 1\,Hz$ and has to be compared to the direct nuclear nuclear dipole coupling, that, in the dense surface arrangement (2.5\r{A}) leads to next neighbour flipping rates of $6.8\,{\rm kHz}$ and is comparable to the mediated one for a nuclear spin distance of $5\,{\rm nm}$. Thus, in the dense bath at large magnetic fields considered here, the decoherence effect arises mainly from the flip-flop interaction of close nuclear spins which is to a good approximation not affected by second order hyperfine processes whereas the decoherence influence of the weakly interacting remote spins, affected by the mediated contribution, can be neglected compared to the first part (despite the fact that it leads to larger differences in the field).\par
\begin{table}[htb]
\centering
\begin{tabular}{|c|c|c|}\hline 
mutual electron spin coupling:&  electron spin - fluorine   & mutual nuclear  \\
 & nuclear spin coupling: & spin fluorine coupling:\\
$c_{el,el}=\left( \dfrac{\mu_0\,\gamma_{el}^2\,\hbar}{4\,\pi}  \right)$  &$c_{el,fl}=\left( \dfrac{\mu_0\,\gamma_{el}\gamma_{fl}\,\hbar}{4\,\pi}  \right)$ &$c_{fl,fl}=\left( \dfrac{\mu_0\,\gamma_{fl}^2\,\hbar}{4\,\pi}  \right)$  \\\hline
  &  &  \\
 $52\,{\rm MHz\,(nm)}^3$ &$74.4\,{\rm kHz}\,({\rm nm})^3$ & $106.3\,{\rm Hz}\,({\rm nm})^3$\\\hline
\end{tabular}
\caption{Magnitude for different coupling constants. Note that the coupling to the NV center corresponds to the electron spin coupling.}
\label{tab_coupl}
\end{table}

Calculating the noise spectrum and the corresponding noise parameters is in general intractable for a large number of surface spins due to exponentially increasing computational resources. As early as in 1962 Klauder and Anderson showed by very general arguments that a Lorentzian noise distribution has to be expected in case of dipolar couplings to a spin bath\,\citep{Sklauder62}. Since then various approaches, mean-field and exact approaches in certain limits, have been invented to infer the noise properties and calculating the decoherence decay rate, ranging from the simple Ornstein-Uhlenbeck model\,\citep{Sdobrovitski09} to cluster expansion methods\,\citep{Syang08, Ssarma06,Smaze08, Syao06}. In here we used the mean-field method described in\,\citep{Ssousa09,Ssousa07} what corresponds to the lowest order of the cluster expansion methods, the so called pair-correlation approximation, corrected by a mean-field broadening using the method of moments\,\citep{Svleck48}. In the pair-correlation approximation one assumes that each of the flipping processes  in the dipolar coupling\,(\ref{ns_1}) is independent of all other processes, i.e. the Hilbert-space is substituted by a pair Hilbert space $(ij)$ wherein ($H_{ij} $ follows directly from eqn.\,(\ref{ns_1}))
\begin{equation}\label{ns_3} H_{\rm int}^{\rm sf}=\sum_{i>j} \tilde{H}_{ij}=\sum_{(ij)} \tilde{H}_{(ij)} \quad \text{with}\,\,[H_{ij},H_{kl}]=0 \,\,\forall ij,kl\,.   \end{equation} 
This leads to a discrete spectrum of $\delta$-peaks at the different pair-induced transition frequencies and can be justified as long as correlations between those pairs can be neglected, that is as long as the evolution time from an initial thermal state is short enough such that $1-\text{exp}(-q\,N_{\rm flip}^2/N)$ is small\,\citep{Syao06} (with $N_{\rm flip}$ the number of flipped bath spins during the considered time, $q$ the number of nearest neighbours and $N$ the total number of bath spins). Higher orders would lead to additional frequency peaks as well as couplings of the already existing peaks, i.e. a finite lifetime broadening what is taken into account by using a mean-field type approach based on the theory of moments. 
From eqn.\,(\ref{ns_1}) and (\ref{h_deph6}) the operator for the effective field on the NV center follows to be 
\begin{equation} \hat{b}=\sum_i \left(\dfrac{\mu_0}{4\,\pi}\,\dfrac{\gamma_{el}\,\gamma_{n}\,\hbar}{r_i^3}  \right)\,(1-3\,\cos^2\theta_i)\,\sigma_i^z \end{equation}  
and the noise spectrum can be calculated by (see section\,\ref{sect_decoh11})
\begin{equation}\label{ns_5} S(\omega)=\int_{-\infty}^\infty\,\mathrm{d}t\,\ex{\hat{b}(t)\,\hat{b}(0)}  \end{equation}
where in good approximation the initial bath states can be assumed to be uncorrelated with the NV center and at room temperature given by $1/2^N  \mathbbm{1}$\,\citep{Sdobrovitski09}.
Following the calculation presented in\,\citep{Ssousa09} this leads to (neglecting static contributions that were also neglected in the decoherence discussion in section\,\ref{sect_decoh11}).
\begin{equation}\label{ns_6} S(\omega)=2\,\pi\,\sum_{i<j}\dfrac{b_{ij}^2\,\Delta_{ij}^2}{b_{ij}^2+\Delta_{ij}^2}\,\,\dfrac{1}{\sqrt{2\,\pi\,\sigma_{ij}^2}}\left( \exp\left[-\frac{(\omega-E_{ij})^2}{2\,\sigma_{ij}^2}  \right]+\exp\left[-\frac{(\omega+E_{ij})^2}{2\,\sigma_{ij}^2}  \right] \right)  \end{equation}
with
\begin{equation}\begin{split} b_{ij}&=\left( \dfrac{\mu_0}{4\,\pi}\,\dfrac{\gamma_{n}^2\,\hbar}{r_{ij}^3}  \right)\,(3\cos^2\theta_{ij}-1)\\
\Delta_{ij}&=  \dfrac{1}{4}\,\left( A_i-A_j \right) \\
A_i&=  2\, \left(\dfrac{\mu_0}{4\,\pi}\,\dfrac{\gamma_{el}\,\gamma_{n}\,\hbar}{r_i^3}  \right)\,(1-3\,\cos^2\theta_i)  \end{split} \end{equation}
and $\sigma_{ij}$ the mean field broadening used in replacing the original delta-peaks by Gaussian peaks leading to the same second moment as the one obtained by the moment theory
\begin{equation}  \sigma_{ij}^2=\dfrac{b_{ij}^2+\Delta_{ij}^2}{4\,\Delta_{ij}^2\,b_{ij}^2}\,\sum_{k\neq i,j} \left(   b_{ik}^2\,A_i^2+b_{jk}^2\,A_j^2 \right)\,.\end{equation}
\begin{figure}[htb]
\begin{centering}
\includegraphics[scale=0.5]{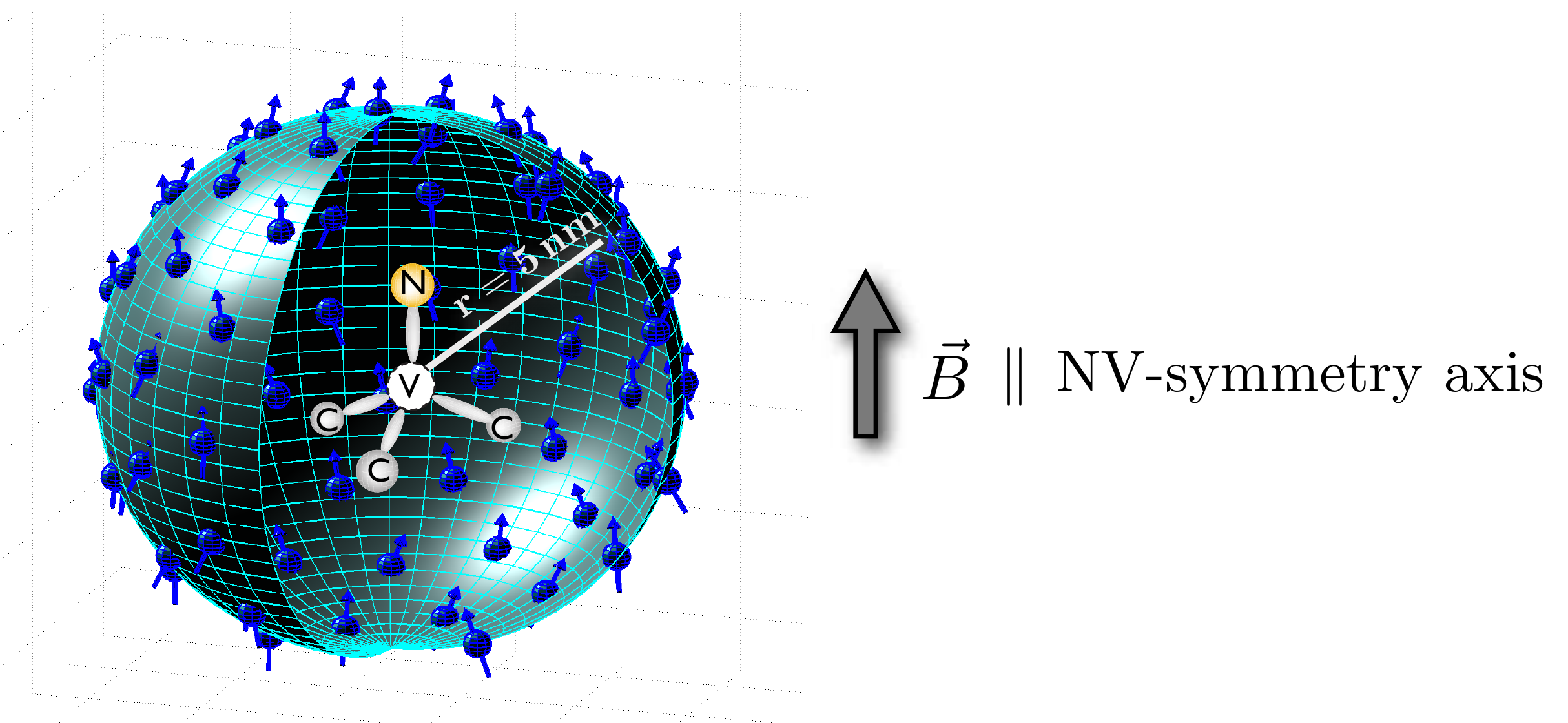}
\caption{\label{bsurfcoupl} Surface spin dephasing influence: The dephasing due to the nuclear surface spins (blue) is modelled by randomly placing nuclear spins on the nanodiamond surface with the diamond assumed to be a sphere of radius $r=5\,{\rm nm}$. The NV center is placed in the center of that sphere with the symmetry axis aligned with the external magnetic field that determines the quantization axis of the surface spins. }
\end{centering}
\end{figure}
The spectrum\,(\ref{ns_6}) was numerically evaluated by randomly distributing nuclear surface spins on a sphere with equal inter-spin distance and placing the NV center in the center of the sphere, i.e. for a sphere radius $r=5\,{\rm nm}$ a number of $5026$ (what corresponds to a distance of $2.5\,$\r{A}) nuclear spins was distributed on the surface (see figure\,\ref{bsurfcoupl}). Assuming an equal quantization direction for as well the NV center and the nuclear spins (corresponding to a strong magnetic field or to a magnetic field parallel to the NV symmetry axis) the noise spectrum was calculated and subsequently fitted to a Lorentzian in order to obtain the noise amplitude $b$ and correlation time $\tau$ defined in\,(\ref{ou_spectrum}). Depending on the number of surface spins an average over different random positions is performed (what however does not lead to a significant change for several thousand spins considered here).\par
We performed the same calculation as well for different numbers of electron surface spins (see figure\,\ref{belspindecoh}) on diamond with radius $5\,{\rm nm}$. In that case the decoherence parameter are very poor, e.g. $T_2=0.1\mu s$, $\tau=21.7\,ns$ and $b=3.6\,{\rm MHz}$ for 30 surface spins and getting worse with further increasing the number of spins. For such a situation decoupling fields in the $\Omega\sim 1\tau\sim {\rm GHz}$ range are required, that, despite still allowing simple two qubit gates (provided the $\Omega$ stability can be achieved), are intractable for multi-qubit applications that require a certain level of individual addressing. Of course it is questionable to what extend the mean-field spectrum calculation, initially based on a pair-approximation, retains its validity in the fast flipping regime of electron spins. To gain some insight in the electron spin case we performed some exact numerical $T_2$ calculations for small electron spin numbers (note that for such small numbers the  results obtained by\,(\ref{ns_6}) are not justified because the spectrum obtained that way differs significantly from a Lorentzian and is very sensitive to the position of the surface spins): For 10 surface spins $T_2$ is already below $0.5\,\mu s$ such that the correlations times obtained by the spectrum approach, even they might be not exact, seem realistic at least in the order of magnitude. Therefore it is crucial to avoid the presence of surface electron spins for performing reliable quantum gates.   

\setcounter{equation}{0}
\section{Dephasing noise simulation and the Ornstein- Uhlenbeck process}\label{sect_decoh}
Dephasing of the nitrogen vacancy center spin states is mainly caused by long range dipolar interactions with a spin-bath, consisting e.g. of substitutional nitrogen atoms (P1 centers) and $^{13} C$ nuclear spins\,\citep{Svandersar12, Slange2010} or in the case of nanodiamonds predominately of unpaired surface spins\,\citep{Stisler09, Sguinness11, Shall10}. Herein a significant energy mismatch of the transition energies prevents flip-flop processes between the vacancy center and the bath spins limiting the influence to a pure dephasing process. However intra-bath flipping processes are not suppressed and occur on the noise correlation timescale $\tau$ such that the NV center is influenced by a random configuration of the bath environment what can be modelled in the mean field approximation as a random magnetic field leading to the frequency shift $b(t)$ and therefore 
\begin{equation} \label{h_deph6} H=b(t)\,S^z\,. \end{equation} 
Now assuming that the back action of the nitrogen spin on the large bath is small and that there is no coherence between different bath spin evolutions, one can model $b(t)$ as a random Gaussian, Markovian and stationary process what is also known as an Ornstein-Uhlenbeck process\,\citep{Slange2010} and can be specified by the two time steady state correlation 
\begin{equation} \label{ou1}   \ex{b(t)\,b(0)}=b^2\,\text{exp}\left(-|t|/\tau  \right)\,  \end{equation}
with $b$ a measure of the coupling strength and $\tau$ the noise correlation time, or alternatively by the expected Lorentzian noise spectrum\,\citep{Sanderson62}
\begin{equation}\label{ou_spectrum} S(\omega)=\int_{-\infty}^\infty\,\mathrm{d}\tau \,\,\ex{b(t)\,b(0)}\,e^{-i\,\omega\tau} = \dfrac{2\,b^2\,\tau}{1+\omega^2\,\tau^2}\,\, .  \end{equation}
 For such a process the evolution of the random field is governed by the Langevin equation\,\citep{Sgillespie95}
\begin{equation}\label{ou2}  \dfrac{\mathrm{d}b(t)}{\mathrm{d}t}=-\dfrac{1}{\tau}\,b(t)+\sqrt{c}\,\Gamma(t)  \end{equation} 
with $\Gamma(t)$ a Gaussian zero-mean noise\,\citep{Sbermudez11} ($\ex{\Gamma(t)}=0$, $\ex{\Gamma(t)\Gamma(t')}=\delta(t-t')$). The parameter $c$ in\,(\ref{ou2}) is related to the definitions in\,(\ref{ou1}) by $b^2=c\,\tau/2$ what follows by solving $\ex{b(t)\,b(0)}$ by means of\,(\ref{ou2}). Exact updating formulas can be obtained for the differential equation\,(\ref{ou2}) and also for the time integral of b(t) given by $y(t+\mathrm{d}t)=y(t)+b(t)\,\mathrm{d}t$\,\citep{Sgillespie95}:
\begin{equation}\label{ou3}\begin{split} b(t+\Delta t)&=b(t)\,\mu+\xi_x\,n_1  \\
			y(t+\Delta t)&=y(t)+b(t)\,\tau\,(1-\mu)+\sqrt{\xi_y^2-\frac{\kappa^2}{\xi_x^2}}\,n_2+\dfrac{\kappa}{\xi_x}\,n_1
 \end{split}\end{equation}
with
\begin{equation}\begin{split}  
\mu&=\text{exp}(-\Delta t/\tau)\\
\xi_x^2 &=(c\,\tau/2)\,(1-\mu^2)\\
\xi_y^2  &=c\,\tau^3\,\left[ \Delta t/\tau-2\,(1-\mu)+1/2\,(1-\mu^2)  \right]\\
\kappa&=(c\,\tau^2/2)\,(1-\mu)^2
    \end{split} \end{equation}
and $n_1$, $n_2$ statistically independent unit normal random numbers. \par
\textbf{Simulation of the noise process: } Simulating the evolution governed by a Hamiltonian of the form\,(\ref{htot_tls}) can be easily performed by numerically solving the Schr\"odinger equation and using the exact updating formulas\,(\ref{ou3}). In here we used a Suzuki-Trotter decomposition splitting the Hamiltonian in the noise $H_n$ and remaining part $H_r$, such  that the time evolution is given by ($\hbar=1$)
\begin{equation}  \exp(-i\,H^{\rm TLS}\,t)=\left[\exp\left(-i\,H_r\, \frac{t}{2n}\right)\,\exp\left(-i\,\int_{\Delta t'=t/n}H_n\,\mathrm{d}t'  \right)\,\exp\left( -i\,H_r\, \frac{t}{2n} \right)   \right]^n+\mathcal{O}\left(\left[\frac{t}{n}\right]^3\right)  \end{equation} 
and the noise part can easily be evaluated using\,(\ref{ou3}).

\setcounter{equation}{0}
\section{Creating an effective $\sigma_z\,\sigma_z$ type coupling}\label{sect_szsz}
A full Ising interaction, that is a $\sigma_z\,\sigma_z$ type coupling in the presence of decoherence is a crucial task towards the way to a universal set of gates, that, once achieved, easily allows to create all other types of interactions by just applying local unitary transformations, straightforward to implement by local microwave pulses on the electron spin transition, e.g. : $\sigma_i^x\,\sigma_j^x=U_x\,\sigma_i^z\,\sigma_j^z\,U_x^\dagger$ or $\sigma_i^y\,\sigma_j^y=U_y\,\sigma_i^z\,\sigma_j^z\,U_y^\dagger$ with $U_x=\text{exp}\left(i\,\pi/4\,(\sigma_y^i+\sigma_y^j)  \right)$ and $U_y=\text{exp}\left(i\,\pi/4\,(\sigma_x^i+\sigma_x^j)  \right)$. \par
Noting that (see also definitions in the main text)
\begin{equation} \label{sising1} H_{zz}=H_{I,\mathcal{M}_1}+H_{I,\mathcal{M}_2}=\sum_{i>j, i\in {\rm neighb}(j)}(J_{i,j}/2)\,\sigma_z^i\sigma_z^j\, ,\end{equation}
allows to use the time-addition methods to implement the $H_{zz}$ time evolution
by either using the Suzuki-Trotter formalism
\begin{equation} \text{exp}\left( -i\,H_{zz}\,t \right) =\left[  R_{\mathcal{M}_1}\left(t/(2n)\right)\,R_{\mathcal{M}_2}\left( t/n \right)\,R_{\mathcal{M}_1}\left(t/(2n) \right)  \right]^n+\mathcal{O}((J_{ij}t/n)^3)\end{equation} or Trotter formalism 
\begin{equation} \text{exp}\left( -i\,H_{zz}\,t \right) =\left[  R_{\mathcal{M}_1}\left(t/n\right)\,R_{\mathcal{M}_2}\left( t/n \right) \right]^n+\mathcal{O}((J_{ij}t/n)^2)\, ,\end{equation}
with $R_{\mathcal{M}_k}(t)=\exp(-i\,H_{I,\mathcal{M}_k}\,t)$.
The switching frequency between the two types of interaction, $\mathcal{M}_1$ and $\mathcal{M}_2$, depends crucially on the Hamiltonian timescales and to avoid a destruction of the decoupling effect, it is important to work in the interaction picture frame of the Hamiltonian $H_{I,\mathcal{M}_k}$, i.e. 
\begin{equation}\begin{split}H_{\mathcal{M}_k}&=H^{0}_{\mathcal{M}_k}+H_{\rm dip}\\  H_{I,\mathcal{M}_k}&=e^{+i\,H_{\mathcal{M}_k}^0\,t}\,H_{\rm dip}\,e^{-i\,H_{\mathcal{M}_k}^0\,t}=\dfrac{1}{2}\,\sum_{i,j}J_{ij}\,S_{\mathcal{M}_k}^{ij} \end{split}\end{equation}
wherein $H_{\mathcal{M}_k}^0=\sum_i \Omega_i/2\,\sigma_i^x$ represents the decoupling field,  $H_{\rm dip}=1/2\,\sum_{i>j i\in\text{neighb}(j)}J_{ij}\,\sigma_z^i\,\sigma_z^j$ the dipole-dipole coupling term and $S_{\mathcal{M}_1}^{ij}=s_+^i\,s_-^j+\text{h.c.}$, $S_{\mathcal{M}_2}^{ij}=s_+^i\,s_+^j+\text{h.c.}$ with $s_\pm$ the corresponding ladder operators in the $\sigma_x$ eigenbasis.\par
Recalling the interaction picture definition, i.e. the connection of the interaction picture state $\ket{\psi(t)}_I$ with the original one $\ket{\psi(t)}$:
\begin{equation}  \ket{\psi(t)}_I=e^{+i\,H_{\mathcal{M}_k}^0\,t} \,\ket{\psi(t)}=e^{+i\,H_{\mathcal{M}_k}^0\,t}\,e^{-i\,H_{\mathcal{M}_k}\,t}\,\ket{\psi(0)}\end{equation}
it is obvious that an interaction picture time evolution can be created by the following pulse sequence:
\begin{equation}\label{eff_int} R_{\mathcal{M}_k}(t):=e^{-i\,H_{\mathcal{I,M}_k}t} =e^{+i\,H_{\mathcal{M}_k}^0\,t}\,e^{-i\,H_{\mathcal{M}_k}\,t}=E_{\mathcal{M}_k}(t)\,U_{\mathcal{M}_k}(t)\,\end{equation}
 with $U_{\mathcal{M}_k}(t)=\text{exp}(-i\,H_{\mathcal{M}_k}\,t)$ and $E_{\mathcal{M}_k}(t)=\text{exp}(i\,H_{\mathcal{M}_k}^0\,t)$. There are essentially three options to create\,(\ref{eff_int}): Directly implementing the pulse sequence, adjusting the total time and making use of the different timescales in the Hamiltonian parts or by using an echo sequence. While the first method is obvious we will discuss the last two approaches in the following. \\
Note also that a larger Trotter slicing parameter $n$ does not necessarily increase the fidelity; the time interval must still be large enough such that $\Omega\,(t/n)>2\pi$ and therefore the off-resonant contributions average to zero and one really ends up with a proper manifold interaction in the interaction picture frame. For the typical parameters considered in this paper $Jij\simeq 26\,{\rm kHz}$ and $\Omega\simeq 1\,{\rm MHz}$ optimal values for $t=\pi/(2J_ij)$ are given by $n\simeq 2,3$.\par

\textbf{Time adjustment: } 
Noting the two different timescales in the Hamiltonian $H_{\mathcal{M}_k}$, i.e. the one of the dipolar coupling $J_{ij}$ in the ${\rm kHz}$ range and the decoupling field strength in the ${\rm MHz}$ range it is possible to create the effective interaction\,(\ref{eff_int}) by merely adjusting the total time. This can be seen by noting that the switching process in the time addition of $\mathcal{M}_1$ and $\mathcal{M}_2$ happens on the timescale of $\sim J_{ij}^{(-1)}$. On this timescale, what forms the typical timescale of $t$ appearing in\,(\ref{eff_int}), the Hamiltonian contribution $H_{\mathcal{M}_k}^0$ can be viewed as performing multiple $2\,\pi$ pulses that are, up to a global phase, not relevant for the interaction. Therefore the pulse $E_{\mathcal{M}_k}(t)$ can be implemented on a much shorter time $t'$ determined by $\Omega^{-1}$ on which to a very good approximation $E_{\mathcal{M}_k}(t')\simeq U_{\mathcal{M}_k}(t')$. $t'$ can be determined by the following consideration (note that global phase contributions are neglected)
\begin{equation}\begin{split} \text{exp}\left(i\,\frac{\Omega}{2}\,\sigma_x\,t  \right)&= \text{exp}\left(\frac{i}{2}\,\sigma_x\,\text{mod}\left[ \Omega\,t,2\,\pi \right]  \right) \\
&=\text{exp}\left(-\frac{i}{2}\,\sigma_x\,(2\,\pi-\text{mod}\left[ \Omega\,t,2\,\pi \right] )  \right)\\
&=\text{exp}\left(-i\frac{\Omega}{2}\,\sigma_x\,\frac{(2\,\pi-\text{mod}\left[ \Omega\,t,2\,\pi \right] )}{\Omega}  \right)\\
&=\text{exp}\left(-i\frac{\Omega}{2}\,\sigma_x\,t' \right)
 \end{split}\end{equation} 
with $t'=(2\,\pi-\text{mod}\left[ \Omega\,t,2\,\pi \right] )/\Omega$.\\
Therefore, if one wants e.g. to implement the Suzuki-Trotter sequence
\begin{equation}  \text{exp}\left( \sum_{\substack{i>j\\i\in \text{neighb}(j)}}\frac{J_{ij}}{2}\,\sigma_i^z\,\sigma_j^z \,t\right)=\left[ R_{\mathcal{M}_1}\left(\frac{t}{2n} \right)\,R_{\mathcal{M}_2}\left(\frac{t}{n} \right)\,R_{\mathcal{M}_1}\left(\frac{t}{2n} \right) \right] +\mathcal{O}\left(\left[ \frac{J_{ij}\,t}{n} \right]^3  \right)\end{equation} 
one can replace $E_{\mathcal{M}_k}(t/(2n))\simeq U_{\mathcal{M}_k}(t_n')$ with $t_n'=(2\,\pi-\text{mod}\left[ \Omega\,t/(2\,n),2\,\pi \right] )/\Omega$ leading to
\begin{equation} R_{\mathcal{M}_k}\left( \frac{t}{2n} \right)\simeq U_{\mathcal{M}_k}\left(  \frac{t}{2n}+t_n'\right)  \end{equation}
what equals to replace $R_{\mathcal{M}_k}\rightarrow U_{\mathcal{M}_k}$ and 
\begin{equation} \dfrac{t}{2\,n}\rightarrow \dfrac{t}{2\,n}+\dfrac{1}{\Omega}\,\left( 2\,\pi - \text{mod}\left[ \frac{\Omega\,t}{2\,n},2\,\pi \right] \right)   \end{equation}
or equivalently the total time $t$ by
\begin{equation} t\rightarrow t+\dfrac{2\,n}{\Omega}\,\left( 2\,\pi - \text{mod}\left[ \frac{\Omega\,t}{2\,n},2\,\pi \right] \right)  \end{equation}
resulting in a simple evolution under the original Hamiltonian $H_{\mathcal{M}_k}$ with an adjusted total gate time.

\textbf{Echo pulses:}
Another possibility of implementing\,(\ref{eff_int}) can be performed eliminating the additional $H_{\mathcal{M}_k}^0$ contribution by echo pulses. Out of\,(\ref{eff_int}) it follows immediately that
\begin{equation} e^{-i\,H_{\mathcal{M}_k}\,t}=e^{-i\,H_{\mathcal{M}_k}^0\,t}\,e^{-i\,H_{I,\mathcal{M}_k}\,t}.  \end{equation}
The goal will be to cancel the additional $H_{\mathcal{M}_k}^0$ contribution by means of  a unitary (echo) $\pi$-pulse $S_\pi$, i.e.
\begin{equation}\begin{split} S_\pi e^{-i\,H_{\mathcal{M}_k}\,t/2} S_\pi^\dagger\,e^{-i\,H_{\mathcal{M}_k}\,t/2}  &=
e^{-i\,[S_\pi\,H_{\mathcal{M}_k}^0\,S_\pi^\dagger]\,t/2}\,e^{-i\,[S_\pi H_{I,\mathcal{M}_k}\,S_\pi^\dagger]\,t/2}\,e^{-i\,H_{\mathcal{M}_k}^0\,t/2}\,e^{-i\,H_{I,\mathcal{M}_k}\,t/2} \\
&\overset{!}{=}e^{-i\,H_{I,\mathcal{M}_k}\,t}
\end{split}\end{equation}
thus leading to the conditions
\begin{equation}\begin{split}     &(i)\quad S_\pi\, H_{I,\mathcal{M}_k}\,S_\pi^\dagger\overset{!}{=}H_{I,\mathcal{M}_k} \\
&(ii)\quad S_\pi\,H_{\mathcal{M}_k}^0\,S_\pi^\dagger\overset{!}{=}-H_{\mathcal{M}_k}^0\\
& (iii)\quad [H_{I,\mathcal{M}_k},H_{\mathcal{M}_k}^0]\overset{!}{=}0\,.
\end{split} \end{equation}
It is straightforward to verify that those conditions are fulfilled by performing a $\pi$-pulse in $\sigma_z$ or $\sigma_y$, i.e.
\begin{equation} S_\pi=\text{exp}\left( -i\frac{\pi}{2}\,\sum_i\sigma_i^z  \right)\quad \text{or}\quad  S_\pi=\text{exp}\left( -i\frac{\pi}{2}\,\sum_i\sigma_i^y  \right)\, .\end{equation}
Hence the interaction\,(\ref{eff_int}) can be implemented using the following sequence
\begin{equation}  R_{\mathcal{M}_k}(t)=S_\pi^\dagger\,U_{\mathcal{M}_k}(t/2)\,S_\pi\,U_{\mathcal{M}_k}(t/2)\,  .\end{equation}
Originating in the non-commutativity of $H_{I,\mathcal{M}_k}$ and $H_{\mathcal{M}_{k'}}^0$ for $k\neq k'$ this echo pulse sequence has to be applied for each $\mathcal{M}_k$ interaction separately and cannot be implemented globally on the total pulse.

\textbf{Changing between $\mathcal{M}_1$ and $\mathcal{M}_2$: }
The addition of the manifold interaction $\mathcal{M}_1$ and $\mathcal{M}_2$ requires to change between the configurations $\Omega_i=\Omega\,\,\forall i$ and $\Omega_i=-\Omega_j\,\,\forall i\in\text{neighb}(j)$. In practice this means that for changing from $\Omega$ to $-\Omega$ the corresponding phase of the microwave coupling has to be changed by $\Delta \phi=\pi$. Alternatively one can also keep the laser configuration fixed and apply additional pulses on the quantum states, i.e.
\begin{equation} S_\pi^{\mathcal{N}^c}\,H_{I,\mathcal{M}_1}\,S_\pi^{{\mathcal{N}^c}\dagger}= H_{I,\mathcal{M}_2}    \end{equation}
with $ S_\pi^{\mathcal{N}^c}=\text{exp}\left( -i\,\pi/2\,\sum_{i,i\in \mathcal{N}^c}\sigma_z^i  \right)$ and $\mathcal{N}^C$ the space of non-neighbouring qubits.

\subsection{Creating a $\sigma_z\,\sigma_z$-operation by purely global operations?}
\begin{figure}[htb]
\begin{centering}
\includegraphics[scale=0.45]{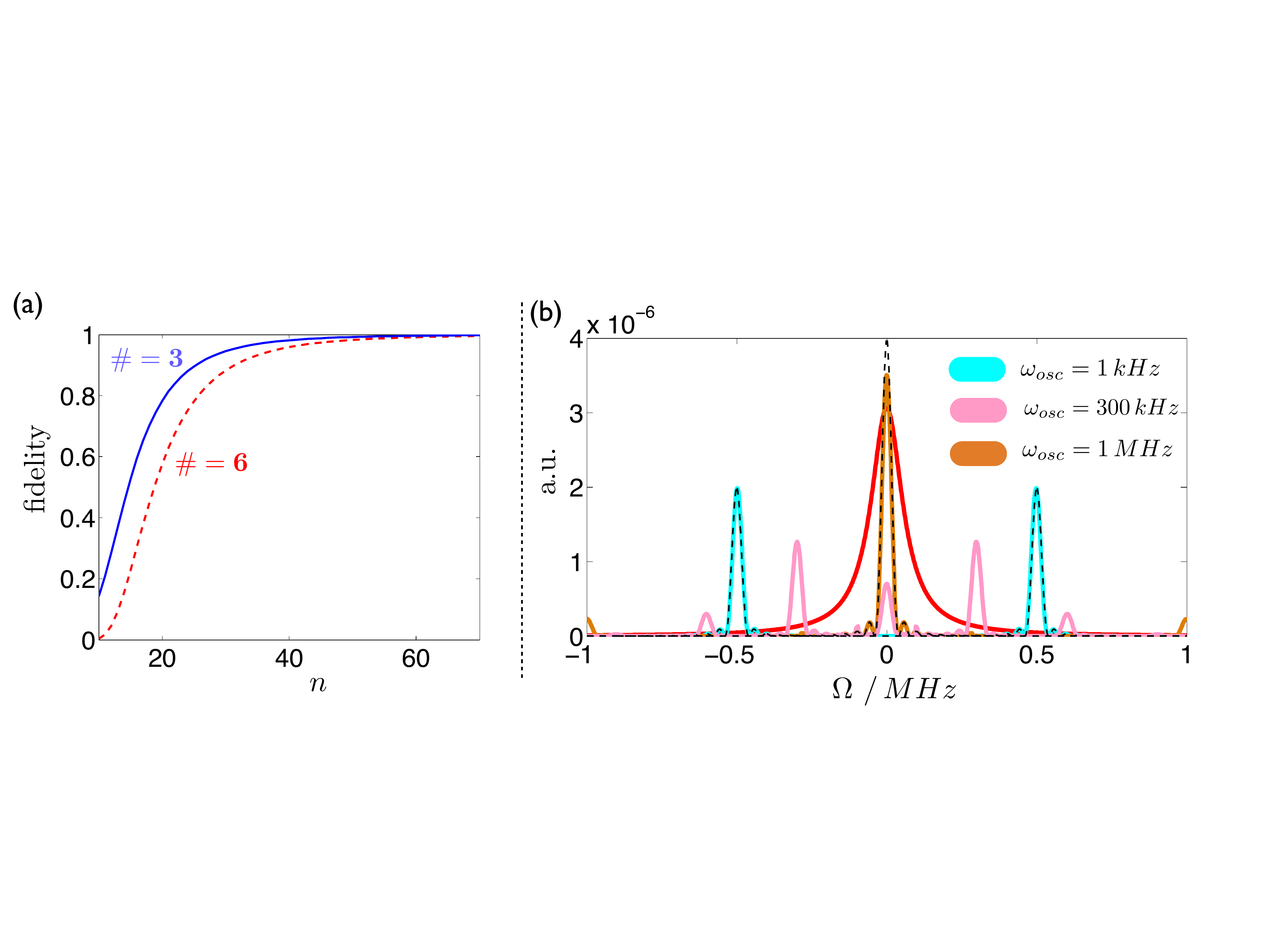}
\caption{\label{bosc_decoupl} \textbf{(a)} Fidelity for creating a $\sigma_z\sigma_z$ -$\pi/2$ pulse for a linear chain of three (blue) and six qubits (red) by directly adding $H_{\mathcal{M}_1}$ and $H_{\mathcal{M}_3}$ according to\,(\ref{sz_dir_add}) with a decoupling field strength $\Omega=1.2\,{\rm MHz}$ and a dipolar coupling $J=26\,{\rm kHz}$. \textbf{(b)}  Oscillating decoupling field in the filter spectrum description for $t=25\,\mu s$. The red curve illustrates the noise spectrum ($T_2=13.3\,\mu s$, $\tau=2.5\,\mu s$, $\sqrt{\ex{b^2}}=30.2\,{\rm kHz}$) and the dashed black lines the filter functions for zero decoupling and $\Omega_0=0.5\,{\rm MHz}$, respectively. The blue, pink and orange lines correspond to filter functions for an oscillating decoupling $\Omega(t)=\Omega_0\,\cos(\omega_{osc}\,t)$ with $\omega_{osc}=1\,{\rm kHz}$,\,\,$\omega_{osc}=300\,{\rm kHz}$ and $\omega_{osc}=1\,{\rm MHz}$, respectively, showing that the decoupling effect vanishes for high oscillation frequencies comparable to the noise correlation time.   }
\end{centering}
\end{figure}

In the attempt for only using global interactions on the NV center array, one might think of the conceptually simpler method of adding up only Hamiltonians with global unique decoupling fields, i.e. the Hamiltonians $H_{\mathcal{M}_1}$ and $H_{\mathcal{M}_3}$
\begin{equation}  H_{\mathcal{M}_k}=\sum_i \dfrac{\Omega_i}{2}\,\sigma_i^x+\dfrac{1}{2}\,\sum_{\substack{i>j\\i\in\text{neighb}(j)}} J_{ij}\,\sigma_i^z\,\sigma_j^z   \end{equation}
with $\mathcal{M}_1$ characterized by $\Omega_i=\Omega\,\,\forall i$ and $\mathcal{M}_3$ by $\Omega_i=-\Omega\,\forall i$. This allows to create a pure $\sigma_z\,\sigma_z$ type coupling noting that
\begin{equation}  H_{\mathcal{M}_1}+H_{\mathcal{M}_3}=\sum_{\substack{i>j\\i\in\text{neighb}(j)}} J_{ij}\,\sigma_i^z\,\sigma_j^z =2\,H_{zz}.  \end{equation}
and using e.g. the Suzuki-Trotter pulse sequence (with $U_{\mathcal{M}_k}=\exp(-i H_{\mathcal{M}_k} t)$)
\begin{equation}\label{sz_dir_add} U_{\mathcal{M}_1}\left( \frac{t}{2n}\right)\,U_{\mathcal{M}_3}\left( \frac{t}{n}\right)\,U_{\mathcal{M}_1}\left( \frac{t}{2n}\right)=e^{-i\,2\,H_{zz}\,t}+\mathcal{O}\left( \left[ \frac{\Omega\,t}{n} \right]^3 \right) \,.  \end{equation}
This concept seems to have several advantages: First it requires only global decoupling microwave fields and second it does not require to work in the interaction picture (beside the original rotating frame with the laser frequency) thus avoiding all the additional complications discussed in the previous section. However there exists a crucial difference in the error scaling which now depends on the the magnitude of the decoupling field instead of the much weaker dipolar coupling. Therefore timesteps such that $\Omega\,t/n <1$ are required and thus $t/n \sim 1/\Omega \sim \tau$, recalling that an effective decoupling is only obtained when the Rabi frequency of the decoupling field is of the order of the inverse noise correlation time (see figure\,\ref{bosc_decoupl}\,a). Such high oscillation frequencies between the decoupling configurations $\mathcal{M}_1$ and $\mathcal{M}_3$ however destroy the decoupling effect as shown in figure\,\ref{bosc_decoupl}\,b with effectively the decoupling field averages to zero for fast flipping frequencies. To conclude, creating a $\sigma_z\sigma_z$ type coupling by the time addition method and at the same time preserving the decoupling effect is only possible in the effective interaction picture Hamiltonian frame and therefore cannot be performed by purely global addressing.

\subsection{Heisenberg Chain simulation}
\begin{figure}[htb]
\begin{centering}
\includegraphics[scale=0.45]{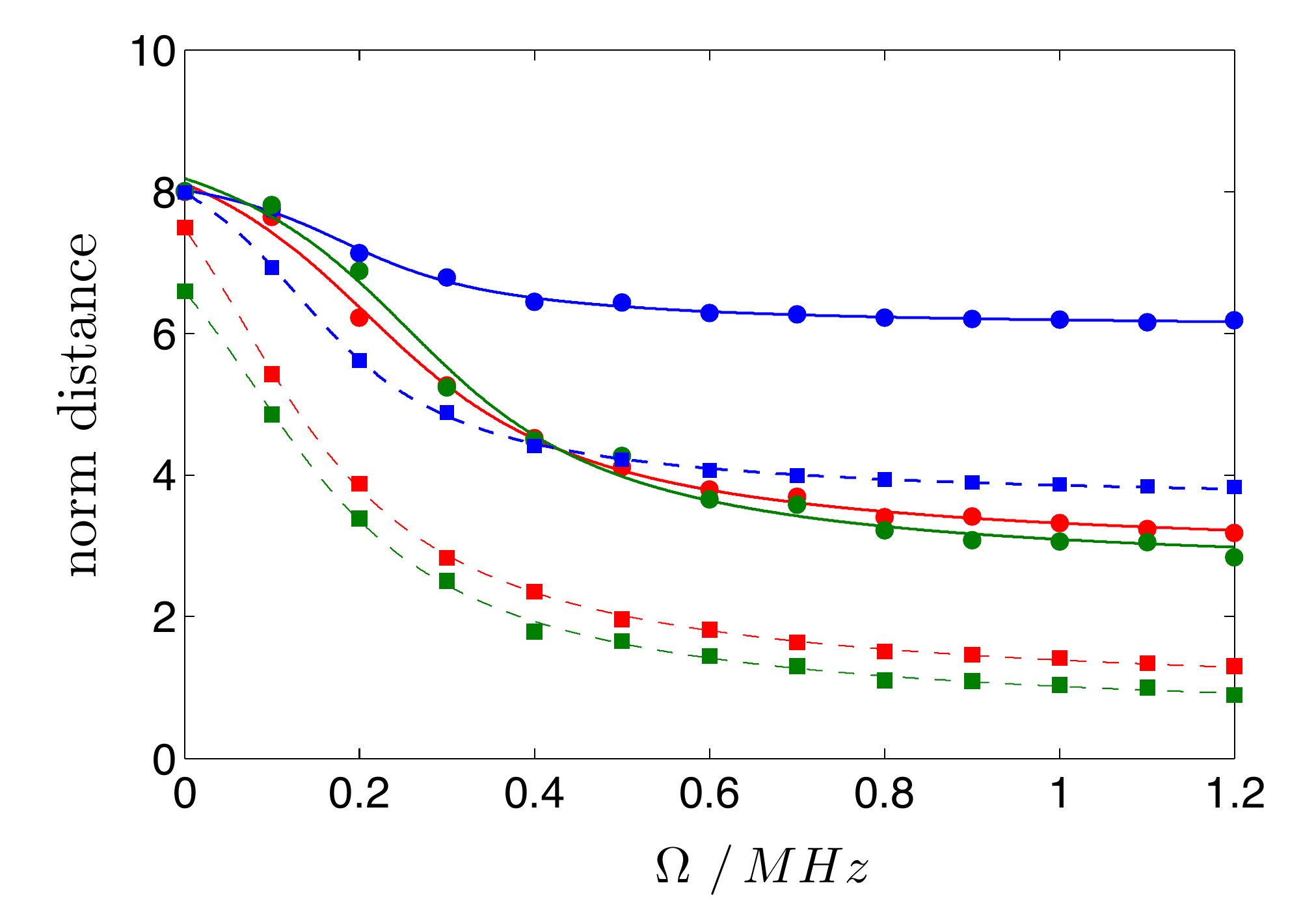}
\caption{\label{b_heisenberg} Heisenberg XXZ-chain with 5 qubits and $\delta=1.5$. The Hilbert-Schmidt norm distance for a time evolution of $t=\pi/(2J)$ is plotted vs the decoupling field and one (blue), two (red) and three (green) Suzuki-Trotter cycles. Dashed lines correspond to the pure $\sigma_z^i\,\sigma_z^j$ contribution. (Hilbert-Schmidt norm distance: $\text{tr}[(U'-U)^\dagger(U'-U)]$ with $U$ the perfect and $U'$ the imperfect realisation.)   }
\end{centering}
\end{figure}

As a second application beside the cluster state creation discussed in the main text, we illustrate the possibility of simulating a Heisenberg-chain Hamiltonian of the form $H=-1/2\,\sum_{j=1,\mu}^N J_\mu\,\sigma_\mu^j\,\sigma_\mu^{j+1}$ with $\mu=\{ x,y,z \}$ and $J_y=J_z=J$, $J_x=\delta\,J$.
This task essentially requires two steps: Creating the individual parts by using the methods described in the previous discussion, namely the $\sigma_x^j\,\sigma_x^{j+1}$-contribution and the trivial $\sigma_y^j\,\sigma_y^{j+1}+\sigma_z^j\,\sigma_z^{j+1}=S_{\mathcal{M}_1}^{j,j+1}$ contributions and in a subsequent step adding those two contribution in time. The accuracy of such a procedure is illustrated in figure\,\ref{b_heisenberg} for a $\theta=\pi/2$ evolution that is limited by non-nearest neighbour interaction as well as the number of Trotter cycles (2-3) such that each time slice $dt$ still provides $\Omega\,dt >2\pi$ to guarantee the desired $H_{\mathcal{M}_k}$ Hamiltonian form.

\setcounter{equation}{0}
\section{Systematic error compensation}

\subsection{Compensation cycle for the two qubit gate interaction}\label{sect_two_comp}
The two qubit gate interaction corresponds essentially to a single qubit rotation in a more complicated two qubit manifold $\mathcal{M}_1$ and $\mathcal{M}_2$, respectively. Introducing the (unknown) systematic error $\epsilon$ this results, assuming $\Omega \gg J$ in the following Hamiltonians
\begin{equation}\label{hm2}\begin{split}   H_{I,\mathcal{M}_1}&\simeq\dfrac{J}{2}(1+\epsilon)\left(s_1^+\,s_2^-+s_1^-\,s_2^+\right)\quad \text{for }\Omega_1=\Omega_2\\
H_{I,\mathcal{M}_2}&\simeq\dfrac{J}{2}(1+\epsilon)\left(s_1^+\,s_2^++s_1^-\,s_2^-\right)\quad \text{for }\Omega_1=-\Omega_2  \end{split}  \end{equation}
what can also be written as
\begin{equation}\label{se2} H_{I,\mathcal{M}_k}\simeq\dfrac{J}{2}(1+\epsilon) \sigma_{x}^{\mathcal{M}_k}    \end{equation}
with $\sigma_x^{\mathcal{M}_k}$ the $\sigma_x$ operation defined in the manifolds $\{ \ket{+-},\ket{-+} \}$ and $\{ \ket{++}, \ket{--} \}$ for k=1 and k=2, respectively (see figure\,3\,(a) in the main text).\par
References\,\citep{Sbrown04, Swimperis94} provide a method to remove the systematic error contribution based on noting that multiples of $2\,\pi$ pulses with respect to the ideal error-less gate end up with pure $\epsilon$ contributions which can, together with the property
\begin{equation}\label{se3}  \sigma_\phi+\sigma_{-\phi}=2\,\cos\phi\,\sigma_x  \end{equation}
wherein $\sigma_\phi=\cos\phi\,\sigma_x+\sin\phi\,\sigma_y$, be used to create a compensation cycle by means of a Suzuki-Trotter time addition of the two contributions in\,(\ref{se3}). Restricting this method to the lowest order of the Suzuki-Trotter expansion (in order to keep the pulse sequence simple and additionally because higher orders require more time and therefore end up in a loss of additional fidelity by decoherence), it follows that for $2\,\pi\,\cos\phi=-\theta/2$
\begin{equation}\label{se4}   M_\phi^{\{\epsilon\}}(\pi)\,M_{3\phi}^{\{\epsilon\}}(2\,\pi)\,M_{\phi}^{\{\epsilon\}}(\pi)\,M^{\{\epsilon\}}(\theta)=M^{\{\epsilon=0\}}(\theta)+\mathcal{O}(\epsilon^3)    \end{equation}
with 
\begin{equation}\label{se5}  M_\phi^{\{\epsilon\}}(\theta)=\text{exp}\left(-i\,\frac{\theta}{2}\,(1+\epsilon)\,\sigma_\phi \right)  \end{equation}
and
\begin{equation}\label{se6} \sigma_\phi=\cos\phi\,\sigma_x+\sin\phi\,\sigma_y\,. \end{equation}
Instead of giving a detailed derivation of the compensation sequence here we refer to reference\,\citep{Sbrown04} and will discuss the similar idea for the multiparticle case in\,\ref{szszcompsect} as well as the special case of the $M_\phi(\pi)\,M_{3\phi}(2\pi)M_\phi(\pi)$ sequence in section\,\ref{multipartfail} with an emphasis on the complication with increasing particle number. Note that the change of the rotation axis $\phi$ can be accomplished by using
\begin{equation}\label{se_dec}  e^{-i\,\theta/2\,\sigma_\phi} = e^{-i\,\phi/2\,\sigma_z}\,e^{-i\,\theta/2\,\sigma_x}\,e^{i\,\phi/2\,\sigma_z}   \end{equation}
what allows to rewrite the rotation $M_\phi$ as
\begin{equation}\label{se8}  M_\phi^{\{ \epsilon\}}(\theta)=T_\phi\,\,M_{\phi=0}^{\{ \epsilon \}}(\theta)\,T_\phi^\dagger  \end{equation}
with $T_\phi=\text{exp}(-i\phi/2\,\sigma_z)$.
\par
\begin{figure}[htb]
\begin{centering}
\includegraphics[scale=0.55]{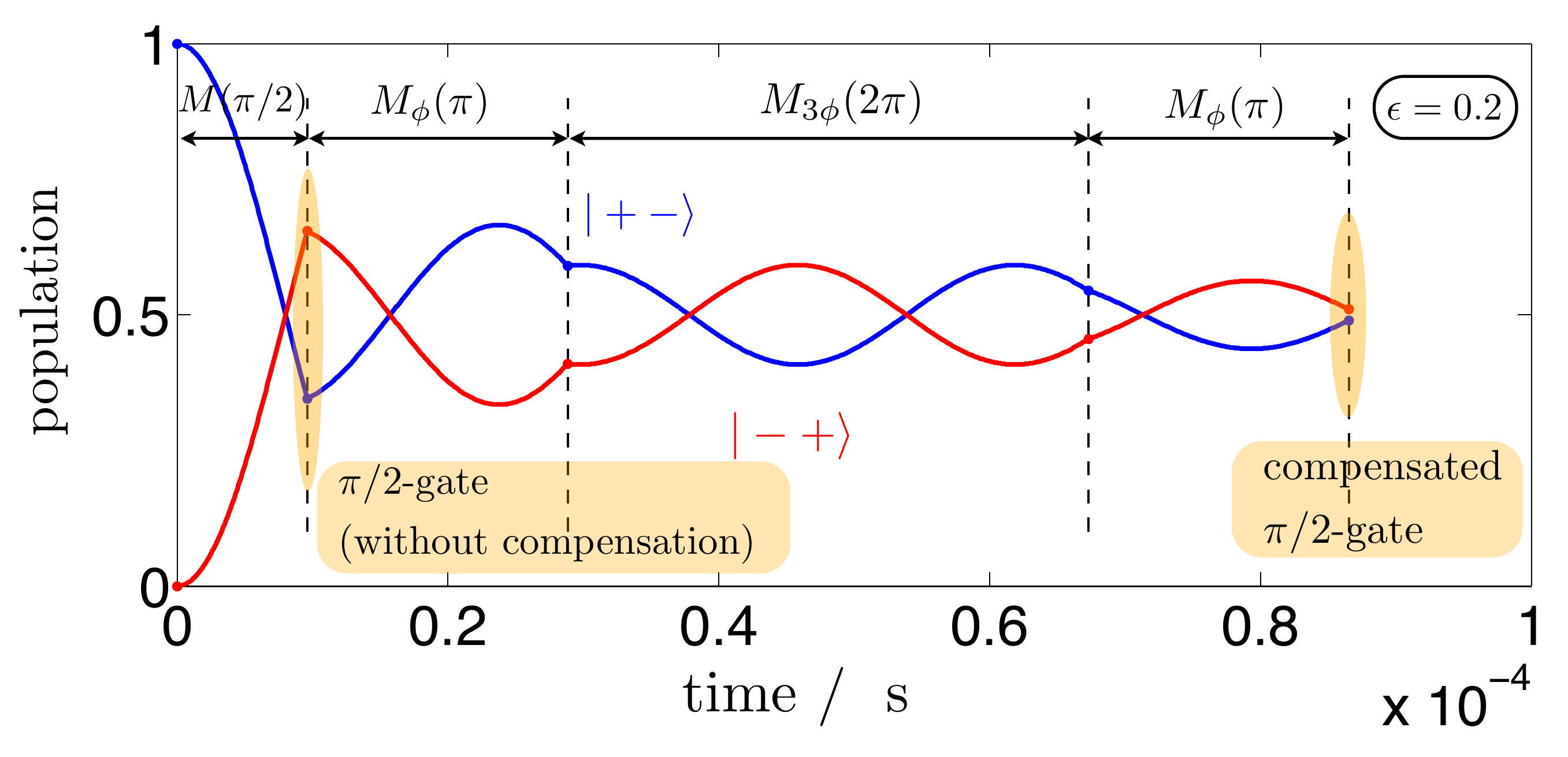}
\caption{\label{b_compcyc} $\pi/2$ gate in the $\mathcal{M}_1$ manifold including systematic error compensation sequence for two qubits. The systematic error is characterized by $\epsilon=0.2$ and the dipolar coupling by $J=26\,{\rm kHz}$. No decoherence effects are taken into account. }
\end{centering}
\end{figure}

The application of\,(\ref{se4}) to the evolution of Hamiltonian\,(\ref{se2}) is straightforward by just replacing $\sigma_\phi, \sigma_x, \sigma_y, \sigma_z$ appearing in\,(\ref{se5}-\ref{se8}) by the manifold counterparts $\sigma_\phi^{\mathcal{M}_k},  \sigma_x^{\mathcal{M}_k}, \sigma_y^{\mathcal{M}_k}, \sigma_z^{\mathcal{M}_k}$. More precise it follows that
\begin{equation} \sigma_z^{\mathcal{M}_1}=\begin{cases}1/2\,(\sigma_x^1-\sigma_x^2) \\ \sigma_x^1\\\sigma_x^2 \end{cases}   \qquad   \sigma_z^{\mathcal{M}_2}=\begin{cases}1/2\,(\sigma_x^1+\sigma_x^2) \\ \sigma_x^1\\\sigma_x^2 \end{cases}  \end{equation}
wherein the last two options have additional contributions that however do not affect the states in the gate manifolds. In summary, a compensated gate is obtained using the sequence (see figure\,\ref{b_compcyc})
\begin{equation} \label{se6b}  M_\phi^{\{\epsilon\}}(\pi)\,M_{3\phi}^{\{\epsilon\}}(2\,\pi)\,M_{\phi}^{\{\epsilon\}}(\pi)\,M^{\{\epsilon\}}(\theta)=M^{\{\epsilon=0\}}(\theta)+\mathcal{O}(\epsilon^3)   \end{equation}
with
\begin{equation} M^{\{ \epsilon \}}(\theta)=\exp\left( -i\,H_{I,\mathcal{M}_k}\,\theta/J \right)  \end{equation}
\begin{equation}  2\,\pi\,\cos\phi=-\theta/2 \end{equation}
and $M_\phi^{\epsilon}(\theta)=T_\phi\,M^{\{ \epsilon \}}(\theta)\,T_\phi^\dagger$ with $T_\phi=\exp(-i\phi/2\,\sigma_z^{\mathcal{M}_k})$.

\subsection{Compensated $\sigma_z\,\sigma_z$ interaction for multiple qubits}\label{szszcompsect}
\begin{figure}[htb]
\begin{centering}
\includegraphics[scale=0.49]{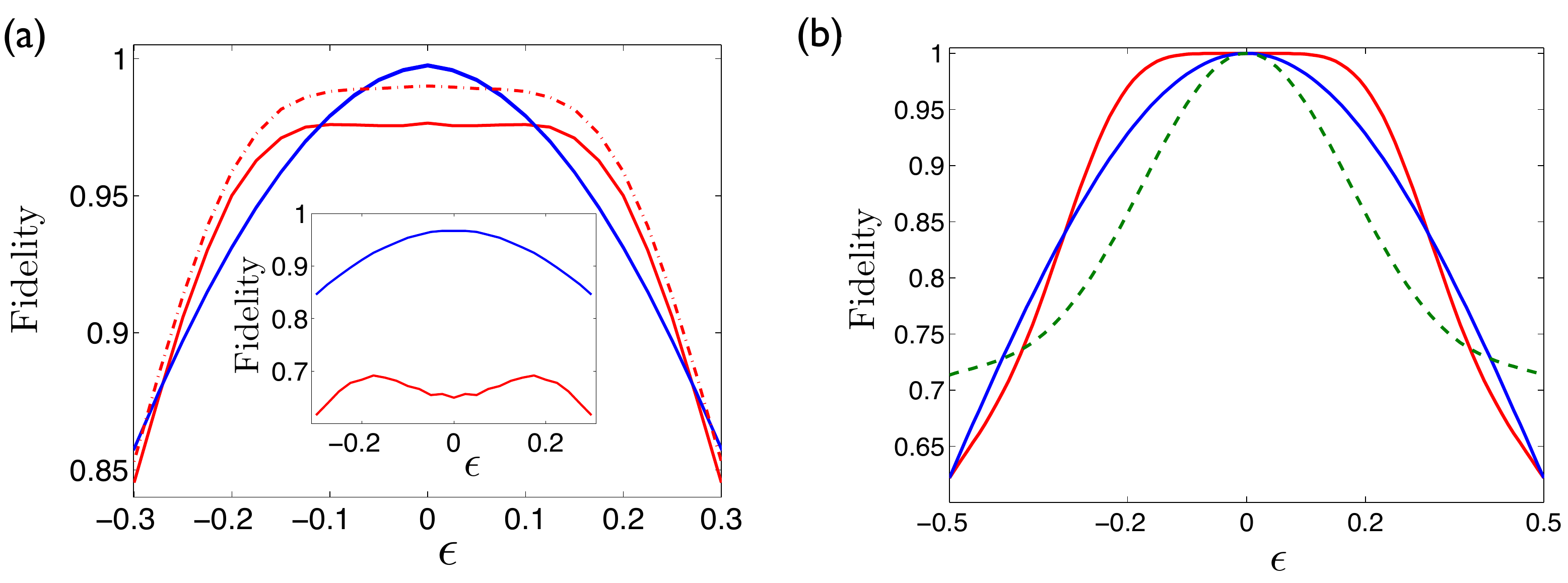}
\caption{\label{b_multierror} \textbf{Multiqubit error compensation} \textbf{(a)} Fidelity vs systematic error $\epsilon$ ($\epsilon_{12}=\epsilon,\epsilon_{23}=-\epsilon,\epsilon_{34}=0.75\epsilon$) for creating a four qubit cluster state without (blue) and with a compensation sequence (red) for $\Omega=5\,{\rm MHz}$ (continuous), $\Omega=10\,{\rm MHz}$ (dashed-dotted) and $\Omega=1\,{\rm MHz}$ (inset) assuming only nearest neighbour interactions (see section\,\ref{sect_nnn}). (Noise parameters: $\tau=2.5\cdot 10^{-6}\,s$, $b=30.2\,{\rm kHz}$, ${T_2}_{\Omega=0}=13.3\,\mu s$. Dipolar Coupling: $J_{ij}=J=26\,{\rm kHz}$.) \textbf{(b)} Fidelity before (blue) and after (red) systematic error compensation\,(\ref{zzcomp5}) for a one dimensional cluster state creation of four qubits and assuming an exact $\sigma_i^z\,\sigma_j^z$ creation (neglecting decoherence). The green dashed lines show the results when applying the advanced\,(\ref{adv1}) sequence, clearly illustrating the failing of that sequence in the multiparticle case.  }
\end{centering}
\end{figure}

Assuming systematic errors $\epsilon_{ij}$ in the dipole dipole coupling term resulting in the Hamiltonian
\begin{equation}\label{hzz2}  H_{zz}^{\{\epsilon\}}=\dfrac{J}{2}\,\sum_{\substack{i>j\\i\in\text{neighb}(j)}}(1+\epsilon_{ij})\,\sigma_i^z\,\sigma_j^z \end{equation}
leads, beside the desired contribution, to additional $\epsilon_{ij}$ dependent terms in the defective evolution operator
\begin{equation}\label{zzcomp1b} M^{\{\epsilon  \}}(\theta)=\text{exp}\left(-i\,\frac{\theta}{2}\,\sum_{i,j} \sigma_z^i\,\sigma_z^j  \right)\,\text{exp}\left(-i\,\frac{\theta}{2}\,\sum_{i,j}\epsilon_{i,j}\,\sigma_z^i\,\sigma_z^j  \right) \,. \end{equation}
Noting that the individual contributions in the sum of the evolution operator commute such that it can be also be written as a product of the individual components, it is advantageous to analyze first simpler situation
\begin{equation}\label{zzcomp1a}  M_{ij}^\epsilon(\theta)= \text{exp}\left(-i\,\frac{\theta}{2}\,(1+\epsilon_{ij})\,\sigma_z^i\,\sigma_z^j  \right)=\text{exp}\left(-i\,\frac{\theta}{2}\,\sigma_z^i\,\sigma_z^j   \right) \,\text{exp}\left(-i\,\frac{\theta}{2}\,\epsilon_{ij}\,\sigma_z^i\,\sigma_z^j   \right) \, . \end{equation}
A contribution only involving the systematic error part can be obtained by any multiple of $2\,\pi$-pulses
\begin{equation} \label{zzcomp1} M_{ij}^\epsilon(n\,2\pi)=(-1)^n\,\text{exp}\left(-i\,n\pi\,\epsilon_{ij}\,\sigma_z^i\,\sigma_z^j   \right)   \end{equation}
with $n\in \mathbb{Z}$. This property forms the basis for creating a compensation sequence provided that the relatively fixed phase (limited by the $2\pi$ condition) can be controlled. The latter is achieved using the properties
\begin{equation}\label{zzcomp2} \left(\sigma_\phi^i+\sigma_{-\phi}^i \right)\,\sigma_z^j= T_\phi\,\left(\sigma_z^i\,\sigma_z^j\right)\,T_\phi^\dagger+ T_{-\phi}\,\left(\sigma_z^i\,\sigma_z^j\right)\,T_{-\phi}^\dagger=2\,\cos\phi\,\sigma_z^i\,\sigma_z^j \end{equation} 
with $T_\phi=\text{exp}(-i\,\phi/2\,\sigma_x^i)$ and $\sigma_\phi=\cos\phi\,\sigma_z-\sin\phi\,\sigma_y$ analogue to\,(\ref{se6}). Performing a Suzuki-Trotter addition of the two components $\sigma_\phi$ and $\sigma_{-\phi}$ therefore leads to (using\,(\ref{zzcomp1}) and (\ref{zzcomp2}))
\begin{equation}\label{zzcomp3}  M_{ij,\phi}^\epsilon(n\,2\,\pi)\,M_{ij,-\phi}^\epsilon(n\,4\,\pi)\,M_{ij,\phi}^\epsilon(n\,2\,\pi)=\text{exp}\left( -i\,n\,4\pi\,\epsilon_{ij}\cos\phi\,\sigma_z^i\,\sigma_z^j  \right)+\mathcal{O}(\epsilon_{ij}^3) \, \end{equation}
wherein $M_{ij,\phi}^\epsilon(\theta)=T_\phi\,M_{ij}^\epsilon(\theta)\,T_\phi^\dagger$.
Comparing this result to the systematic error contribution in\,(\ref{zzcomp1a}) it is obvious that the error part can be compensated by\,(\ref{zzcomp3}) if
\begin{equation} 4\,n\,\pi\,\cos\phi=-\frac{\theta}{2}\,  \end{equation}
in which case a corrected gate operation is obtained using the sequence
\begin{equation}\label{zzcomp4} M_{ij,\phi}^\epsilon(n\,2\,\pi)\,M_{ij,-\phi}^\epsilon(n\,4\,\pi)\,M_{ij,\phi}^\epsilon(n\,2\,\pi)\,M_{ij}^\epsilon(\theta)=\text{exp}\left(-i\,\frac{\theta}{2}\,\sigma_z^i\,\sigma_z^j   \right)+\mathcal{O}(\epsilon_{ij}^3)\, .  \end{equation}
Taking into account that the error scales as $\mathcal{O}(n\,\epsilon_{ij}^3)$ the parameter $n$ should be as small as possible, leading to the obvious choice $n=1$. It is straightforward to extend the compensation analysis to the complete evolution\,(\ref{zzcomp1b}) what finally leads to 
\begin{equation}\label{zzcomp5}   M_\phi^{\{\epsilon\}}(2\,\pi) M_{-\phi}^{\{\epsilon\}}(4\,\pi)\,M_\phi^{\{\epsilon\}}(2\,\pi)\,M^{\{\epsilon\}}(\theta)=M^{\{\epsilon=0\}}(\theta)+\mathcal{O}(\epsilon_{ij}^3) \end{equation} 
with $M_\phi(\theta)=T_\phi\,M(\theta)\,T_\phi^\dagger$, $8\,\pi\,\cos\phi=-\theta$ and $T_\phi=\text{exp}\left( -i\,\phi/2\,\sum_{i|i\in \mathcal{N}^c}\sigma_i^x \right)$ and $\mathcal{N}^c$ the space of non-neighbouring qubits. As an example the method is illustrated for the creation of a four qubit cluster state in figure\,\ref{b_multierror} (a), showing that high decoupling fields are required in presence of decoherence in order to benefit from the compensation despite the significant increased gate time that is problematic in terms of decoherence. 

\subsection{Failing of the advanced method for the multiparticle case }\label{multipartfail}
As indicated in\,(\ref{zzcomp4}) smaller values for $n$ lead to an improved scaling of the compensation method, i.e. to a smaller remaining error contribution and in the case when decoherence is relevant, to a significant reduction of the compensated gate time. Thus one could ask if also $n=1/2$ would be possible. Following the general setup it is obvious that for this choice relation\,(\ref{zzcomp1}) is not valid any more and therefore $\epsilon$ dependent contributions appear in the compensation sequences\,(\ref{zzcomp4}) and\,(\ref{zzcomp5}). However for the case of a single particle rotation, and so also for the compensated two qubit gate discussion in section\,\ref{sect_two_comp}, there exists exactly such a compensation sequence given by\,(\ref{se4}), (\ref{se6b})
\begin{equation} \label{adv1} M_\phi^{\{ \epsilon \}}(\pi)\,M_{3\phi}^{\{ \epsilon \}}(2\,\pi)\,M_\phi^{\{ \epsilon \}}(\pi)\end{equation}
such that for $2\,\pi\,\cos\phi=-\theta/2$
\begin{equation}\label{adv211} M_\phi^{\{ \epsilon \}}(\pi)\,M_{3\phi}^{\{ \epsilon \}}(2\,\pi)\,M_\phi^{\{ \epsilon \}}(\pi)\,M^{\{ \epsilon  \}}(\theta)= M^{\{ \epsilon=0  \}}(\theta)+\mathcal{O}(\epsilon^3)\,.  \end{equation}
Note that, as expected, the choice of rotation angles $\phi$ differs from the one in\,(\ref{zzcomp4}). Before analyzing the failing of this advanced scheme for the multiparticle case it is useful to consider first its working mechanism for a single particle. In that case $M^{\{ \epsilon \}}(\theta)=\text{exp}\left( -i\,\theta/2\,\sigma_z \right)$ and $M_\phi^{\{ \epsilon \}}=T_\phi\,M^{\{ \epsilon \}}\,T_\phi^\dagger$ with $T_\phi=\exp(-i\,\phi/2\,\sigma_x)$ and accordingly
\begin{equation}  M_\phi^{\{ \epsilon \}}(\pi)\,M_{3\phi}^{\{ \epsilon \}}(2\,\pi)\,M_\phi^{\{ \epsilon \}}(\pi)=e^{-i\,\pi/2\,\epsilon\,\sigma_\phi}\,\left[e^{-i\,\pi/2\,\sigma_\phi}\,e^{-i\,\pi\,\epsilon\,\sigma_{3\phi}}\,e^{-i\,\pi/2\,\sigma_\phi}\right]\,e^{-i\,\pi/2\,\epsilon\,\sigma_\phi} \,. \end{equation}
Note that not only $\epsilon$ dependent contributions survive due to the fact that  two of the contributions in\,(\ref{adv1}) are not multiples of $2\,\pi$. From\,(\ref{zzcomp2}), (\ref{zzcomp4}) and the Suzuki-Trotter expansion it turns out that a valid compensation sequence can be constructed if the term in brackets fulfils the following relation
\begin{equation} \label{adv5} e^{-i\,\pi/2\,\sigma_\phi}\,e^{-i\,\pi\,\epsilon\,\sigma_{3\phi}}\,e^{-i\,\pi/2\,\sigma_\phi}=e^{-i\,\pi\,\epsilon\,\sigma_{-\phi}}   \end{equation}
such that
\begin{equation}\begin{split}   M_\phi^{\{ \epsilon \}}(\pi)\,M_\phi^{\{ \epsilon \}}(2\,\pi)\,M_\phi^{\{ \epsilon \}}(\pi)&= e^{-i\,\pi/2\,\epsilon\sigma_\phi}\,e^{-i\,\pi\,\epsilon\,\sigma_{-\phi}}\, e^{-i\,\pi/2\,\epsilon\,\sigma_\phi}\\
	&=\text{exp}\left( -i\,2\,\pi\cos\phi\,\epsilon\,\sigma_z \right)\,.
\end{split}\end{equation}
It therefore remains to prove the validity of\,(\ref{adv5}). Using relation\,(\ref{se_dec}) allows to express it as
\begin{equation}\label{adv6}\begin{split}  e^{-i\,\pi/2\,\sigma_\phi}\,&e^{-i\,\pi\,\epsilon\,\sigma_{3\phi}}\,e^{-i\,\pi/2\,\sigma_\phi}   = e^{-i\,\pi/2\,\sigma_\phi}\,e^{-i2\phi\,\sigma_x}\,e^{-i\,\pi\,\epsilon \sigma_{-\phi}}\,e^{i\,2\phi\,\sigma_x}    e^{-i\,\pi/2\,\sigma_\phi}  \\
&=\left(e^{i\phi\,\sigma_x}\,e^{-i\,\pi/2\,\sigma_\phi}\,e^{-i\phi\,\sigma_x}  \right)\,e^{-i\,\pi\epsilon\,\sigma_{-\phi}}\,\left(e^{i\phi\,\sigma_x}\,e^{-i\,\pi/2\,\sigma_\phi}\,e^{-i\phi\,\sigma_x}  \right)\\
&=e^{-i\,\pi/2\,\sigma_{-\phi}}\,e^{-i\,\pi\,\epsilon\,\sigma_{-\phi}}\,e^{-i\,\pi/2\,\sigma_{-\phi}}=e^{-i\,\pi\,\epsilon\,\sigma_{-\phi}}
 \end{split}\end{equation}
where again condition\,(\ref{se_dec}) was used together with the crucial requirement that
\begin{equation}\label{adv7}  e^{-i\,\pi/2\,\sigma_\phi}\,e^{-i\phi\,\sigma_x}=e^{+i\phi\,\sigma_x}\, e^{-i\,\pi/2\,\sigma_\phi}\end{equation}  
what follows directly by noting that $\text{exp}\left(-i\,\pi/2\,\sigma_\phi  \right)=-i\,\sigma_\phi$ and $\{ \sigma_\phi,\sigma_x \}=0$. \par
Now let us return to the case of multiple particles. Herein the same concepts can be applied as in section\,\ref{szszcompsect} despite now using the pulse sequence\,(\ref{adv1}). However it turns out that in such a situation the condition analogue to\,(\ref{adv7})
\begin{equation}\label{adv17} \left[T_\phi\,e^{-i\pi/2\,\sum_{i>j|i\in \text{neighb}(j)}\sigma_z^i\,\sigma_z^j}\,T_\phi^\dagger\right]\,e^{-i\,\phi\,\sum_{i\in\mathcal{N}^c}\sigma_x^i}\underset{i.g.}{\neq}   e^{+i\,\phi\,\sum_{i\in\mathcal{N}^c}\sigma_x^i}\, \left[T_\phi\,e^{-i\pi/2\,\sum_{i>j|i\in \text{neighb}(j)}\sigma_z^i\,\sigma_z^j}\,T_\phi^\dagger\right] \end{equation}
with $T_\phi$ and $\mathcal{N}^c$ defined as below eqn.\,(\ref{zzcomp5}), is not fulfilled. That this is in general not valid can best be seen by considering the example of three particles in a linear chain in which case
\begin{equation} e^{-i\,\pi/2\,\left[ \sigma_z^1\,\sigma_\phi^2+\sigma_\phi^2\,\sigma_z^3 \right]}\,e^{-i\,\phi\,\sigma_x^2} = e^{-i\,\phi\,\sigma_x^2}\, e^{-i\,\pi/2\,\left[ \sigma_z^1\,\sigma_\phi^2+\sigma_\phi^2\,\sigma_z^3 \right]} \end{equation}
what reveals that due to the double commutation with two $\sigma_\phi$ contributions there is no net sign change in the phase when commuting $\text{exp}\left( -i\,\phi\,\sigma_x^2 \right)$ in contrast to the situation for a single particle in\,(\ref{adv7}). That prevents the application of relation\,(\ref{se_dec}) and hence $\sigma_\phi$ terms cannot be cast in $\sigma_{-\phi}$ as in\,(\ref{adv6}). One should note that this issue can still be fixed for the three particle case by choosing $T_\phi$ in\,(\ref{adv17}) such that the phase change is performed on the first and third particle in what case the left and right hand side of that equation would be equal thus allowing to  use a compensation cycle of the form\,(\ref{adv211}). For more than three particles such a possibility however does not exist. \\
To conclude: Due to an even number of permutations in\,(\ref{adv17}) (i.g. two permutations for a linear chain, four for a two dimensional array and six for a three dimensional one) it is in general not possible to extend the $n=1/2$ compensation sequence\,(\ref{adv211}) to the multiparticle case (see figure\,\ref{b_multierror}(b)). Therefore the general concept\,(\ref{zzcomp5}) has to be used in that situation, despite being less efficient in both time and $\epsilon$ compensation range.

\subsection{Non-next nearest neighbour couplings in the compensation sequence}\label{sect_nnn}
The Hamiltonians\,(\ref{hm2}) and\,(\ref{hzz2}) used to describe the compensation mechanism have been defined up to next-nearest neighbours, i.e. higher order contributions were neglected so far. However, whereas those contributions give only a small correction for simple $\pi/2$ multiqubit pulses, as can be seen from the numerical results in the main text, they are crucial in the compensation sequence. Due to the long pulse times of such a sequence higher order contributions cannot be neglected. One has to distinguish between two types of higher order couplings: Even couplings defined as next-nearest neighbour and third, fifth, \dots nearest neighbour couplings and odd couplings as second, fourth, \dots nearest and diagonal couplings. Whereas the first ones have always the form of a $\sigma_z\sigma_z$ coupling and are automatically compensated for by the next-nearest neighbour based compensation sequence, the second class consists just of the reduced manifold $\mathcal{M}_k$ couplings and is not removed by the compensation sequence. This failure for odd couplings is as well originating from the fact that the compensation pulse acts on both qubits involved such that\,(\ref{zzcomp2}) is not fulfilled any more, as on the fact that the manifold interaction does not commute with the $\sigma_z\sigma_z$ type contributions such that a splitting of the evolution in a perfect and defective part as in\,(\ref{zzcomp1b}) is not possible any more.\par
\begin{figure}[htb]
\begin{centering}
\includegraphics[scale=0.4]{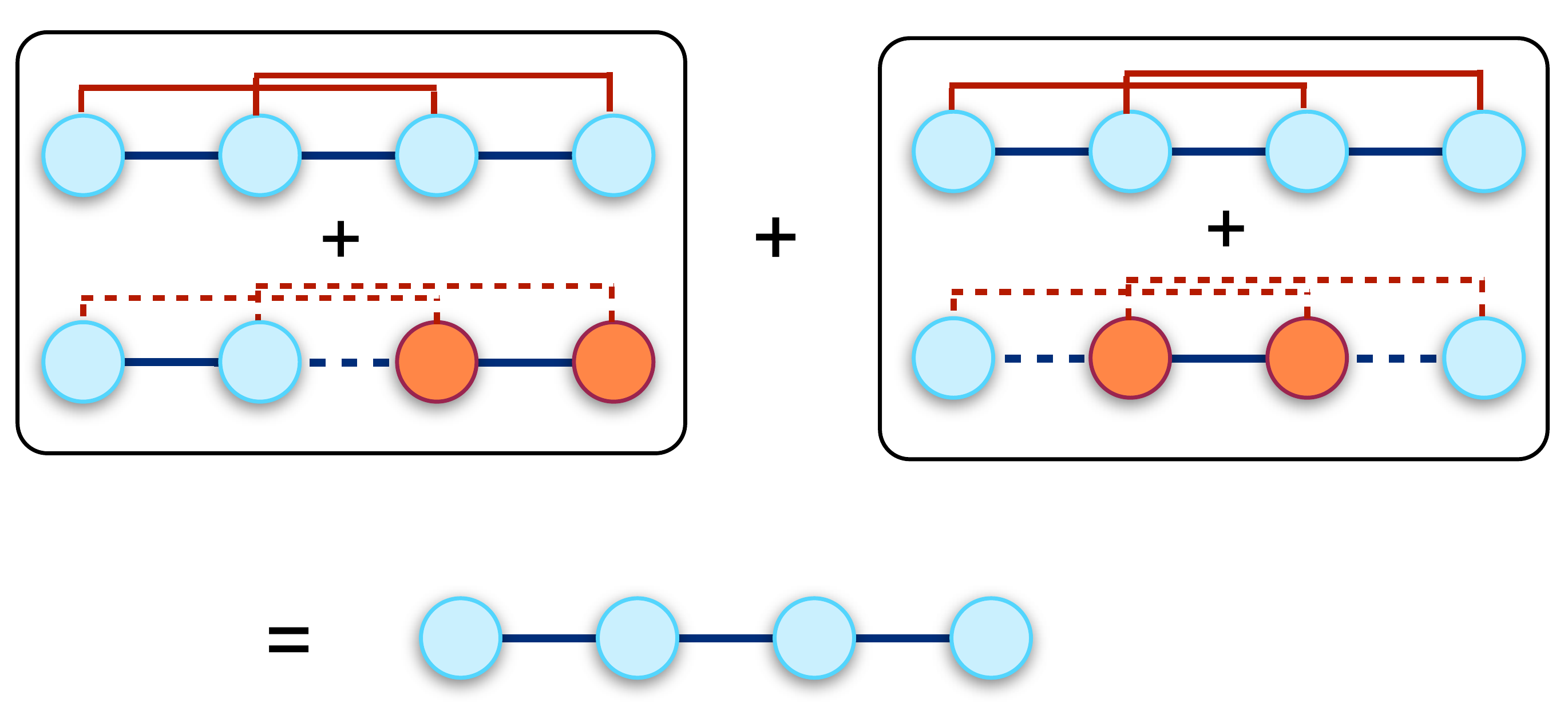}
\caption{\label{b_multinnnelim} Removing 2nd nearest neighbour interactions for a linear four qubit configuration. Each of the blocks removes 2nd nearest neighbour couplings; however both blocks are needed to restore the proper next nearest neighbour interaction. Blue lines denote $\sigma_z\sigma_z$ and red lines $\mathcal{M}_1$ interactions and dashed lines denote a negative sign of the corresponding interaction. Red marked qubits denote that the interaction is embedded between a $U$ and $U^\dagger$ pulse with $U=\text{exp}(-i\pi/2\sigma_x)$. Couplings higher than second order are neglected in the illustration.   }
\end{centering}
\end{figure}

A solution to that problem consists of eliminating odd order couplings from the beginning by adding different contributions in time. Examples for that are given for a two qubit state in figure\,4\,(b) in the main text and in figure\,\ref{b_multinnnelim} for second nearest neighbour couplings. With the evolutions obtained that way as a starting point higher orders can be removed in a concatenated way taking benefit of the different evolution timescales dependent on the qubit distance. For the multiqubit compensation sequence\,(\ref{zzcomp4}) it is sufficient to remove odd order couplings up to the second order (e.g. up to couplings of qubit one to five), because higher orders effects are too weak for giving significant contributions.

\end{document}